\pdfoutput=1
\documentclass[floatfix]{emulateapj}
 \usepackage{amsmath}
\usepackage[boxed]{algorithm2e}
\usepackage{algorithmic}
\usepackage{enumerate}
\usepackage{epsfig,subfigure,color,amsmath,graphicx,multirow,subfigmat,ctable,bm}
\usepackage{natbib}

\newcommand{\beq}{\begin{equation}}
\newcommand{\eeq}{\end{equation}}
\newcommand{\beqs}{\begin{eqnarray}}
\newcommand{\eeqs}{\end{eqnarray}}

\input{epsf} 
\newcommand{\bfi}{\begin{figure} \epsfxsize=8cm \epsffile}
\newcommand{\efi}{\end{figure}}

\begin{document}
\title{Testing the ABS method with the simulated Planck temperature maps}
\author{Jian Yao$^{1}$, Le Zhang$^{1,6}$,  Yuxi Zhao$^{1,2}$, Pengjie Zhang$^{1,3,4,6}$, Larissa Santos$^{5}$, Jun Zhang$^{1,6}$ \\
{~}\\
$^{1}$ Department of Astronomy, Shanghai Jiao Tong University, Shanghai, 200240, China\\
$^{2}$ Physical Science Division, University of Chicago, Chicago, IL 60637, USA \\
$^{3}$ IFSA Collaborative Innovation Center, Shanghai Jiao Tong University, Shanghai 200240, China\\
$^{4}$ Tsung-Dao Lee Institute, Shanghai 200240, China\\
$^{5}$ CAS Key Laboratory for Researches in Galaxies and Cosmology, Department of Astronomy, University of Science and Technology of China, Chinese Academy of Sciences\\
$^{6}$ Shanghai Key Laboratory for Particle Physics and Cosmology }
\thanks{Email: lezhang@sjtu.edu.cn}
\begin{abstract}

In this study, we apply the Analytical method of Blind Separation (ABS) of the cosmic microwave background (CMB) from foregrounds to estimate the CMB temperature power spectrum from multi-frequency microwave maps.  We test the robustness of the ABS estimator and assess the accuracy of the power spectrum recovery by using realistic simulations based on the seven-frequency Planck data, including various frequency-dependent and spatially-varying foreground components (synchrotron, free-free, thermal dust and anomalous microwave emission), as well as an uncorrelated Gaussian-distributed instrumental noise. Considering no prior information about the foregrounds, the ABS estimator can analytically recover the CMB power spectrum over almost all scales with less than $0.5\%$ error for maps where the Galactic plane region ($|b|<10^{\circ}$) is masked out. To further test the flexibility and effectiveness of the ABS approach in a variety of situations, we apply the ABS to the simulated Planck maps in three cases: (1) without any mask, (2) imposing a two-times-stronger synchrotron emission and (3) including only the Galactic plane region ($|b|<10^{\circ}$) in the analysis. In such extreme cases, the ABS approach can still provide an unbiased estimate of band powers at the level of 1 $\mu\rm{K}^2$ on average over all $\ell$ range, and the recovered powers are consistent with the input values within 1-$\sigma$ for most $\ell$ bins. 
\end{abstract}


\maketitle

\section{Introduction}
The cosmic microwave background (CMB) is one of the most powerful cosmological probes to study the physical processes that occurred in the early universe. The power spectra of its temperature and polarization anisotropies encode detailed information on the statistics of the primordial perturbations, the existence of gravity waves and the physical components of the universe. The accurate measurement of the CMB power spectrum is thus of prime importance in cosmological parameter estimation. For this reason, tremendous experimental efforts such as Boomerang~\citep{2000Natur.404..955D}, MAXIMA~\citep{2002ApJ...568...38H}, DASI~\citep{2002ApJ...568...38H}, VSA~\citep{2003MNRAS.341.1057W}, CBI~\citep{2003ApJ...591..540M}, ACBAR~\citep{2004ApJ...600...32K}, SPT~\citep{2013JCAP...10..060S},  ACT~\citep{2014JCAP...04..014D}, POLARBEAR~\citep{2014ApJ...794..171P}, SPTpol~\citep{2015ApJ...807..151K}, and especially the WMAP~\citep{2003ApJS..148....1B,2013ApJS..208...19H} and Planck satellites~\citep{2014A&A...571A..16P} have already successfully provided measurements of the CMB power spectra, yielding tight constraints on the cosmological parameters. Moreover, the primordial B-modes can probe horizon-scale primordial gravitational waves and play a major role in understanding the inflationary epoch~\citep{zal+seljak:1997,kamionkowski97,Hu2002}. Recently, a number of experiments have been or are being deployed with the goal of accurately measuring the CMB polarization signal, such as QUBIC~\citep{battistelli/etal:2011}, BICEP2~\citep{2014PhRvL.112x1101B}, Ali-CPT~\citep{2017arXiv171003047L}, CORE~\citep{2017arXiv170604516D}, LiteBIRD~\citep{2014JLTP..176..733M}, EPIC/CMBpol~\citep{2009arXiv0906.1188B}, PIXIE~\citep{2011JCAP...07..025K}, PRISM~\citep{2014JCAP...02..006A}.

The accuracy of the CMB power spectra measurements however is limited by several astrophysical foreground radiations present in the sub-millimeter range, mainly including synchrotron, free-free and dust emissions, which originate from within our Galaxy, and the Sunyaev-Zel'dovich effects from extragalactic sources ~\citep{2016A&A...594A..10P}. In reality, when observing the microwave sky, one measures not only the CMB signal itself but a linear mixture of the CMB with other foreground components in addition to the instrumental noise. Thus, it is crucial to separate the cosmological signal from the observed dirty sky so as to recover all the valuable information encoded in the CMB anisotropies.        

Regarding the component separation problem, a great deal of work has been carried out in the literature. Several algorithms, referred to as ``non-blind', which require a prior knowledge of the components frequency dependence, have been dedicated to perform foreground removal. The most commonly used techniques exploiting this method are Wiener Filtering (WF;~\citealt{1994ApJ...432L..75B,1996MNRAS.281.1297T,1999MNRAS.302..663B}) and the Maximum Entropy Method (MEM;~\citealt{1998MNRAS.300....1H}).~\cite{2009ApJS..180..265G} which reconstruct all components based on a Markhov-Chain-Monte-Carlo (MCMC) approach. A joint component separation and the CMB power spectra estimation using the Gibbs sampling approach have been implemented by ~\cite{2004ApJ...609....1J,2004PhRvD..70h3511W,2004ApJS..155..227E,2007ApJ...656..653L,2016A&A...594A..10P}, which is based on a parametric model of the sky components and provides a joint component separation and the Bayesian parameter estimation. On the other hand, the frequency dependence of foreground components is generally poorly known and its uncertainty may lead to unwanted foreground residuals in the cleaned maps. With this motivation, several ``blind'' approaches have been proposed, which make no prior assumption about foregrounds, such as the methods using Independent Component Analysis (ICA;~\citealt{2004MNRAS.354...55B}) and Correlated Component Analysis (CCA;~\citealt{2006MNRAS.373..271B}). The so-called Internal Linear Combination (ILC) approach and its various variants have been extensively applied to the CMB signal processing for foreground removal and to the CMB power spectra estimation~\citep{1996MNRAS.281.1297T,2003PhRvD..68l3523T,2003ApJS..148...97B,2006ApJ...645L..89S,2009A&A...493..835D}. ~\cite{2011MNRAS.418..467R} recently proposed a generalised ILC approach by constructing a multidimensional ILC filter in a needlet domain.

Subsequently, with a procedure based on internal template fitting, ~\cite{2008A&A...491..597L,2012MNRAS.420.2162F} have successfully applied this method to Planck simulations and to WMAP polarization data. In addition, with the assumption of spectral diversity of the various components,~\cite{2003MNRAS.346.1089D,2003MNRAS.345.1101M,2007MNRAS.376..739A,2008arXiv0803.1814C} propose a blind source separation method, the Spectral Matching Independent Component Analysis (SMICA), which has been successfully employed on the Planck data.

Unlike all the above methods which involve heavy computation, ~\cite{2016arXiv160803707Z} recently presented an Analytical method of Blind Separation of the CMB from foregrounds (ABS). Based on the measured cross band power between different frequency bands, the CMB band power spectra can be derived {\it analytically}, which does not rely on any assumption about the foreground components while avoiding multiple parameter fitting. Similar to SMICA, the ABS approach works directly on the cross power spectra. But it uses analytical formula to solve for the CMB power spectrum. It also shares similarities with ILC. In ideal case of no instrument noise,  the ABS output is identical to the power spectrum directly measured from the ILC reconstructed CMB map~\citep{2008A&A...487..775V,2008PhRvD..78b3003S}. In realistic situations where instrument noise exists, the ILC differs from the ABS~\citep{2016arXiv160803707Z}. ABS is designed with the primary goal of achieving unbiased CMB power spectrum measurement. The key ingredient to fulfill this goal is the introduced shift parameter, and the associated convergence test and null test.

Here, we report the first test on the ABS method from simulated Planck observations. For the purpose of assessing the validity of the ABS estimator, we keep the simulated foregrounds as realistic as possible. As the first test, we apply the ABS approach to temperature maps only, considering various sky cuts. It is important to point out that the complicated beam effects are not taken into account in our simulations. Since ~\cite{2016arXiv160803707Z} has already provided a complete description of the mathematical formalism and numerical techniques that in principle can be applied to polarization maps and account for frequency- and position-dependent beams, sky cuts and other non-ideal effects, we propose to dedicate a future paper to systematically test the ABS approach against real-world observations.


The paper is organized as follows. In Sect.~\ref{sect:abs}, we briefly review the ABS approach. In Sect.~\ref{sect:map} we describe our simulated maps and present the application of the ABS to the simulated skies and estimate the accuracy of the CMB temperature power spectrum recovery in Sect.~\ref{sect:test}. Finally, we draw our conclusions in Sect.~\ref{sect:con}.

We will use thermodynamic units throughout this paper, corresponding to a constant CMB power spectrum across frequencies.

\section{The ABS approach}\label{sect:abs}
The ABS approach blindly and analytically subtracts the foregrounds and recover the CMB signal in the spherical harmonic domain rather than in the pixel domain. This approach works on the cross band power spectrum of two frequency maps. The measured data in the multipole bin $\ell$ can be written as
\beq\label{eq:data}
\mathcal{D}^{\rm obs}_{ij}(\ell) = f_if_j\mathcal{D}^{\rm cmb}(\ell) + \mathcal{D}^{\rm fore}_{ij}(\ell) + \delta \mathcal{D}_{ij}^{\rm noise}(\ell)\,.
\eeq
Here, $\mathcal{D}^{\rm obs}_{ij}(\ell)$ denotes the cross band power spectrum of measured temperature maps at the $i$- and $j$-th frequency channels, where $i,j = 1,2\cdots N_f$ and $N_f$ is the total number of frequency channels. In matrix notation, $\mathcal{D}_{ij}(\ell)$ also refers to the $(i,j)$-th entry of the $N_f\times N_f$ matrix $\mathcal{D}(\ell)$. $\mathcal{D}_{ij}^{\rm fore}$ is the cross band power matrix of foregrounds. Here $f_i = 1$ for all channels in the units of thermodynamic temperature. Therefore, $\mathcal{D}^{\rm cmb}$ represents the CMB temperature power spectrum that does not vary with frequency.

The measured cross power spectrum is certainly contaminated by instrumental noise, so that we introduce the noise term, $\delta \mathcal{D}_{ij}^{\rm noise}$, which represents the fluctuations of the instrumental noise in the measurements. The ensemble average of the instrumental noise has been implicitly subtracted out beforehand as it cannot bias the estimate of the CMB power spectrum. In this study, we assume that the instrument noise is an uncorrelated Gaussian distribution, with zero mean and rms levels of $\sigma_{\mathcal{D},i}^{\rm noise}$ for the $i$-th frequency channel, $i=1,2\cdots N_f$. The residual noise hence has the following properties,
\begin{eqnarray}\label{eq:noise}
&&\left<\delta \mathcal{D}_{ij}^{\rm noise}\right> =0\,,\nonumber \\
&&\left<(\delta \mathcal{D}_{ij}^{\rm noise})^2\right> = \frac{1}{2} \sigma_{\mathcal{D},i}^{\rm noise} \sigma_{\mathcal{D},j}^{\rm noise} (1+\delta_{ij})\,.
\end{eqnarray}

\subsection{In the case with no instrumental noise}
Let us first consider the ideal case of no instrumental noise, which motivates the framework of the ABS approach. ~\cite{2016arXiv160803707Z} proves that the CMB power spectrum, $\mathcal{D}^{\rm cmb}$, can be analytically derived through
\beq\label{eq:abs}
\mathcal{D}^{\rm cmb} = \left( \sum_{\mu=1}^{M+1} G^2_{\mu}\lambda_{\mu}^{-1}\right)^{-1}\,, {\rm with}\,\,G_\mu = {\bf f\cdot E}^\mu
\eeq
as long as $M< N_f$, where $M\equiv rank(\mathcal{D}_{ij}^{\rm fore})$ and the order of $\mathcal{D}^{\rm obs}_{ij}(\ell)$ is $N_f$, and the vector ${\bf f} = (f_1,\ldots, f_{N_f})^T$. ${\bf E}^\mu$ and $\lambda_\mu$ stand for the $\mu$-th eigenvector and associated eigenvalue of $\mathcal{D}^{\rm obs}_{ij}(\ell)$. We adopt the normalization condition for eigenvectors, ${\bf E}^\mu \cdot {\bf E}^{\nu} =\delta_{\mu \nu}$. In our calculation, all eigenvalues and the corresponding eigenvectors are ranked in decreasing order of eigenvalues. Note that the rank $M$ depends on the number of independent foreground components. Previously, \cite{2008MNRAS.388..247D} illustrated that the full sky foreground could be well described to a very high accuracy using only a small number of principal foreground components (e.g., $M= 3$).

\subsection{In the case with instrumental noise}\label{sect:noise}
In reality, observations always suffer from instrumental noise. To account for noise, the original Eq.~\ref{eq:abs} can be used to recover the CMB power spectrum with the following modifications. 
\beq\label{eq:abs1}
\mathcal{\hat{D}}^{\rm cmb} = \left( \sum^{\tilde{\lambda}_{\mu}\geq \lambda_{\rm cut}} \tilde{G}^2_{\mu}\tilde{\lambda}_{\mu}^{-1}\right)^{-1} - \mathcal{S}\,.
\eeq
Here, we have introduced new variables, defined by
\begin{eqnarray}\label{eq:noiseD}
&&\mathcal{\tilde{D}}^{\rm obs}_{ij}\equiv \frac{\mathcal{D}^{\rm obs}_{ij}}{\sqrt{\sigma_{\mathcal{D},i}^{\rm noise}\sigma_{\mathcal{D},j}^{\rm noise}}} + \tilde{f}_i\tilde{f}_j\mathcal{S}\,,\nonumber\\
&&\tilde{f_i} \equiv \frac{f_i}{\sqrt{\sigma_{\mathcal{D},i}^{\rm noise}}}\,,\quad
\tilde{G}_{\mu}\equiv {\bf \tilde{f}}\cdot {\bf \tilde{E}}^\mu\,,
\end{eqnarray}
where ${\bf \tilde{E}}^\mu$ and $\tilde{\lambda}_\mu$ are the $\mu$-th eigenvector and corresponding eigenvalue of $\mathcal{\tilde{D}}^{\rm obs}_{ij}$, respectively. According to Eq.~\ref{eq:noise}, the dispersion of each diagonal and off-diagonal elements of $\mathcal{\tilde{D}}^{\rm obs}_{ij}$ in such normalization is $1$ and $1/\sqrt{2}$, respectively. The instrumental noise thus can lead to unphysical (i.e.,noise dominated) eigenmodes with eigenvalues of $|\tilde{\lambda}_\mu| \lesssim1$ in $\mathcal{\tilde{D}}^{\rm obs}_{ij}$, detailed in~\citealt{2016arXiv160803707Z}. For this reason, the threshold $\tilde{\lambda}_{\rm cut}$ in Eq.~\ref{eq:abs1} is not arbitrary but instead has a value of $\sim1$, since all unphysical eigenmodes induced by instrumental noise should be excluded. According to our intensive tests, we find that the estimate of the CMB power spectrum is not sensitive to the threshold in the range of $1/2\lesssim\tilde{\lambda}_{\rm cut}\lesssim1$ . As indicated by the analysis of our simulations, $\tilde{\lambda}_{\rm cut}$ is chosen to be $1/2$ for the best recovery of the CMB.

In Eq.~\ref{eq:abs1}, we also have introduced a useful free parameter $\mathcal{S}$, which shifts the amplitude of the input CMB power spectrum from $\mathcal{D}^{\rm cmb}$ to $\mathcal{D}^{\rm cmb} +\mathcal{S}$. In practice, when solving for $\mathcal{D}^{\rm cmb}$ numerically, we find that a positive {\it shift} parameter is responsible for stabilizing the computation, therefore providing better numerical properties. More importantly, since the ABS method in the absence of $\mathcal{S}$ always returns a positive value (as both $\tilde{G}^2_{\mu}$ and $\tilde{\lambda}_{\mu}$ in Eq.~\ref{eq:abs1} are greater than zero), it should {\it fail} the null test. However, one can avoid such overestimate on the CMB signal by introducing $\mathcal{S}$. This ``shift'' strategy performed successfully when the underlying truth value of $\mathcal{D}^{\rm cmb}$ is much less than the noise level and when it approaches zero. Such strategy becomes notably important for an unbiased determination of  the CMB B-mode power spectrum which has an unknown but close to zero amplitude. In practice, our simulations show that the $\mathcal{S}\sim 50 \sigma_{\mathcal{D}}^{\rm noise}$ is an appropriate value to guarantee a stable computation and a successful null test result (See Appendix in details).

\section{Simulated Planck maps}\label{sect:map}

\begin{figure*}[htpb]
\centering
\mbox{
 \subfigure{
   \includegraphics[width=2.3in] {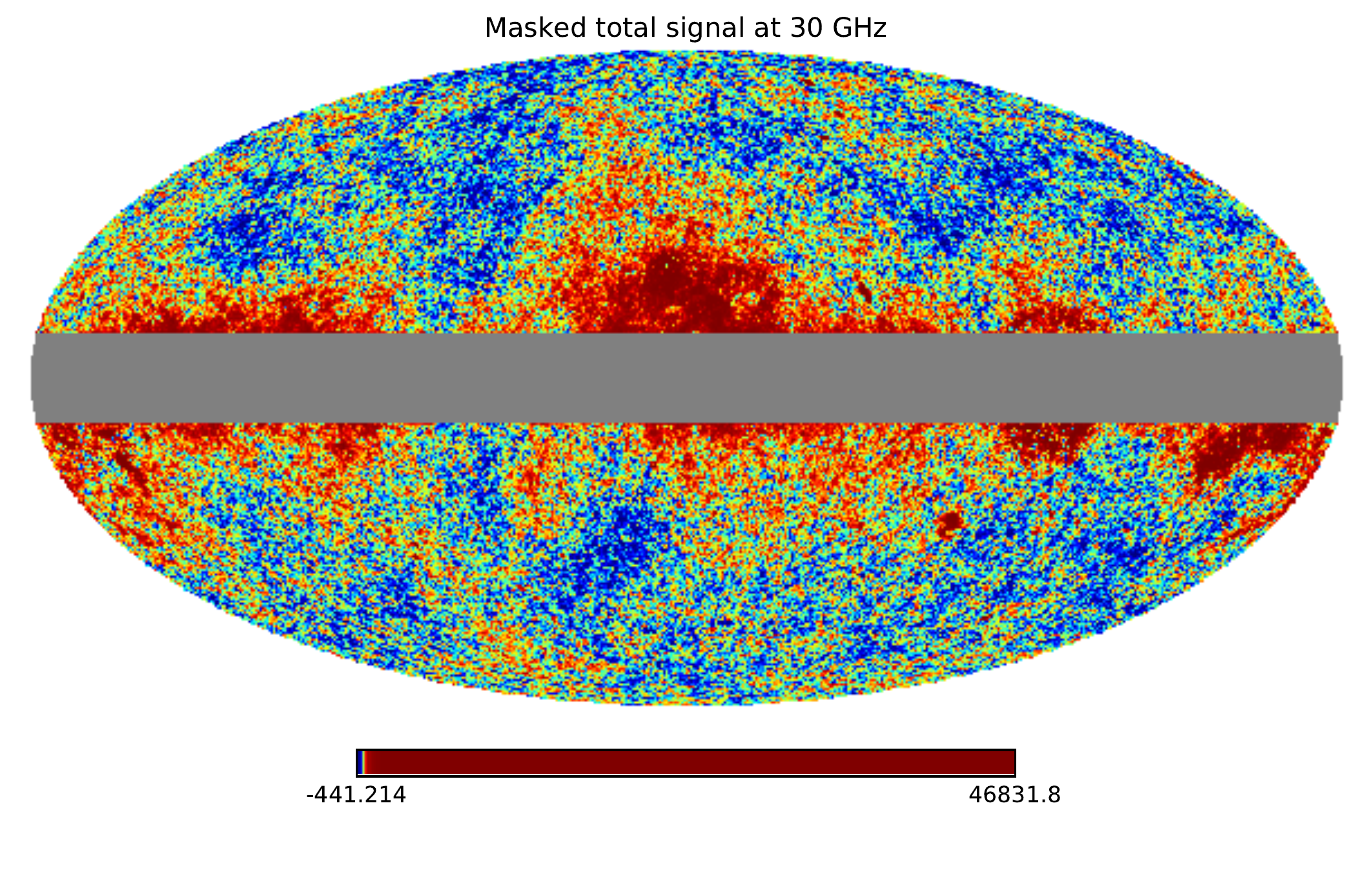}
   }

 \subfigure{
   \includegraphics[width=2.3in] {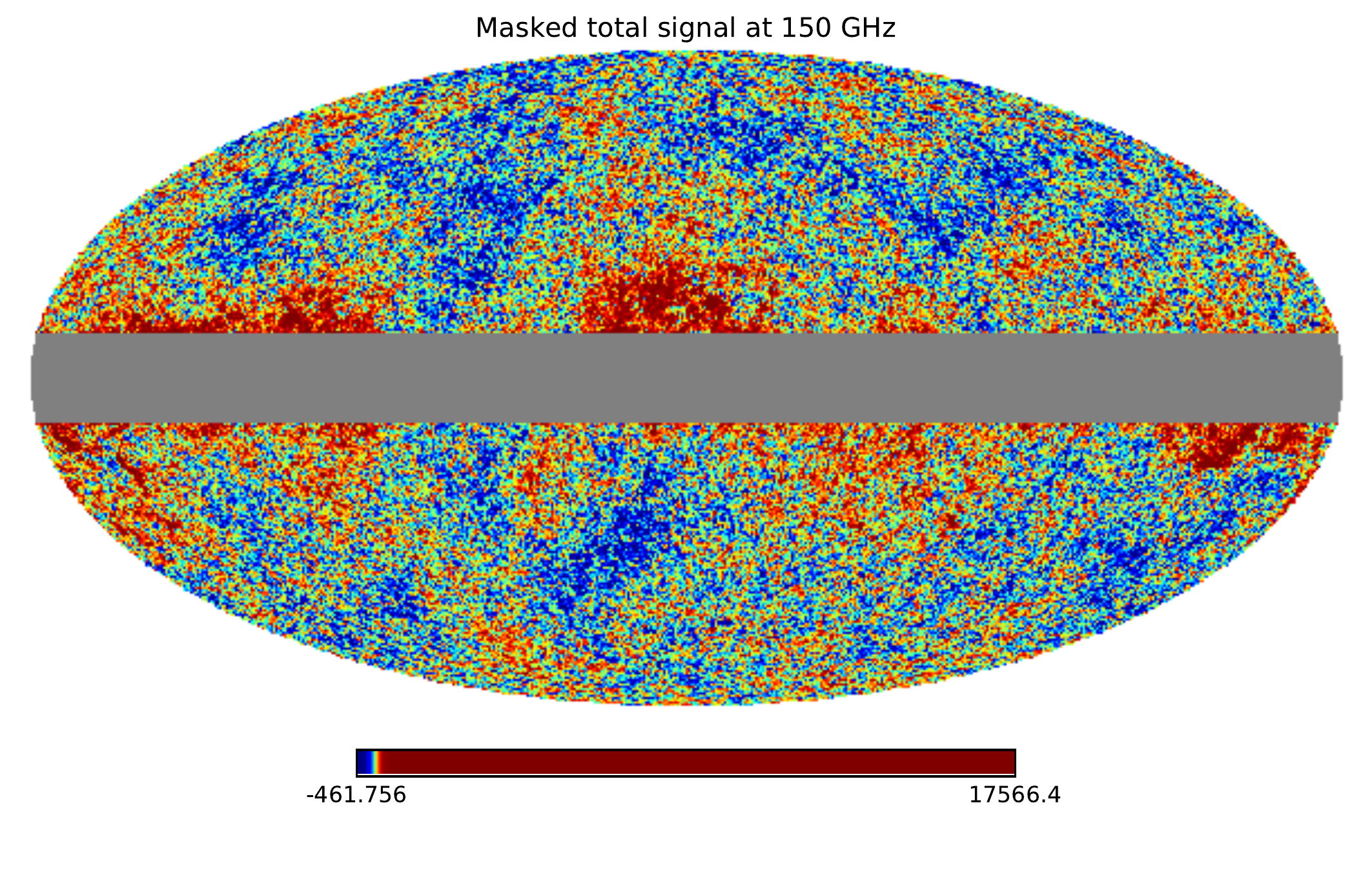}
   }

 \subfigure{
   \includegraphics[width=2.3in] {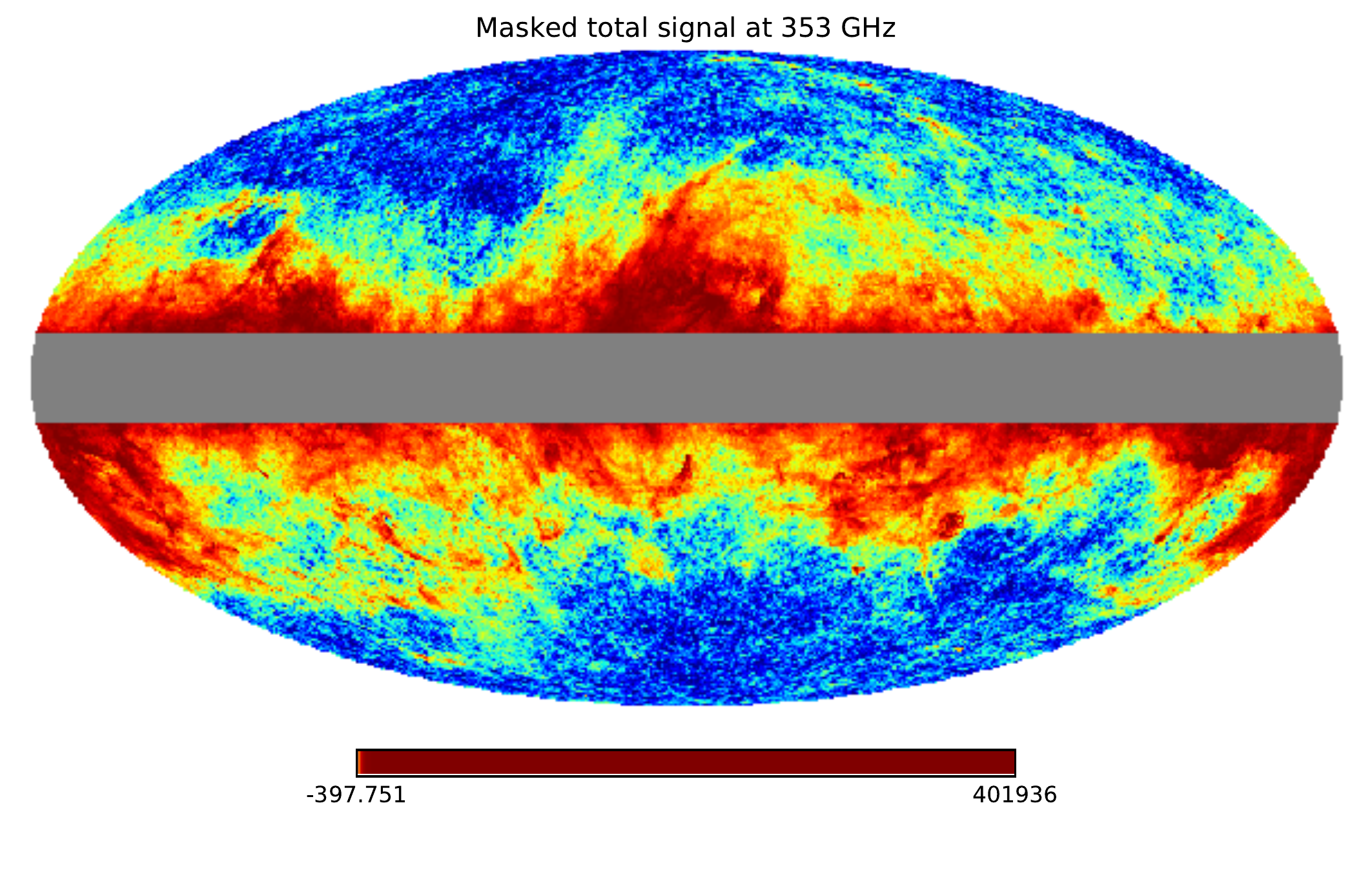}
   }

}

\mbox{
 \subfigure{
   \includegraphics[width=2.3in] {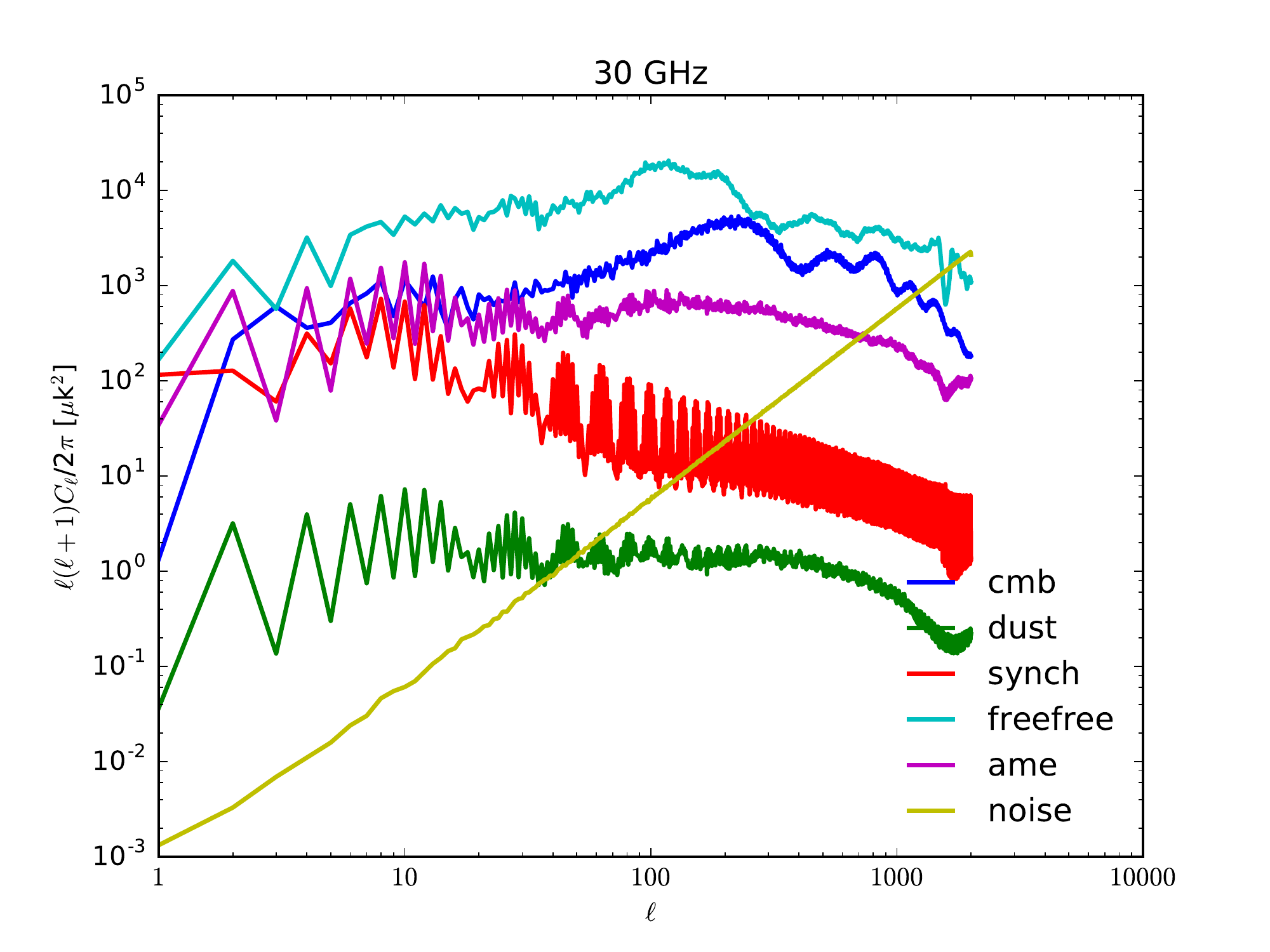}
    }

\subfigure{
   \includegraphics[width=2.3in] {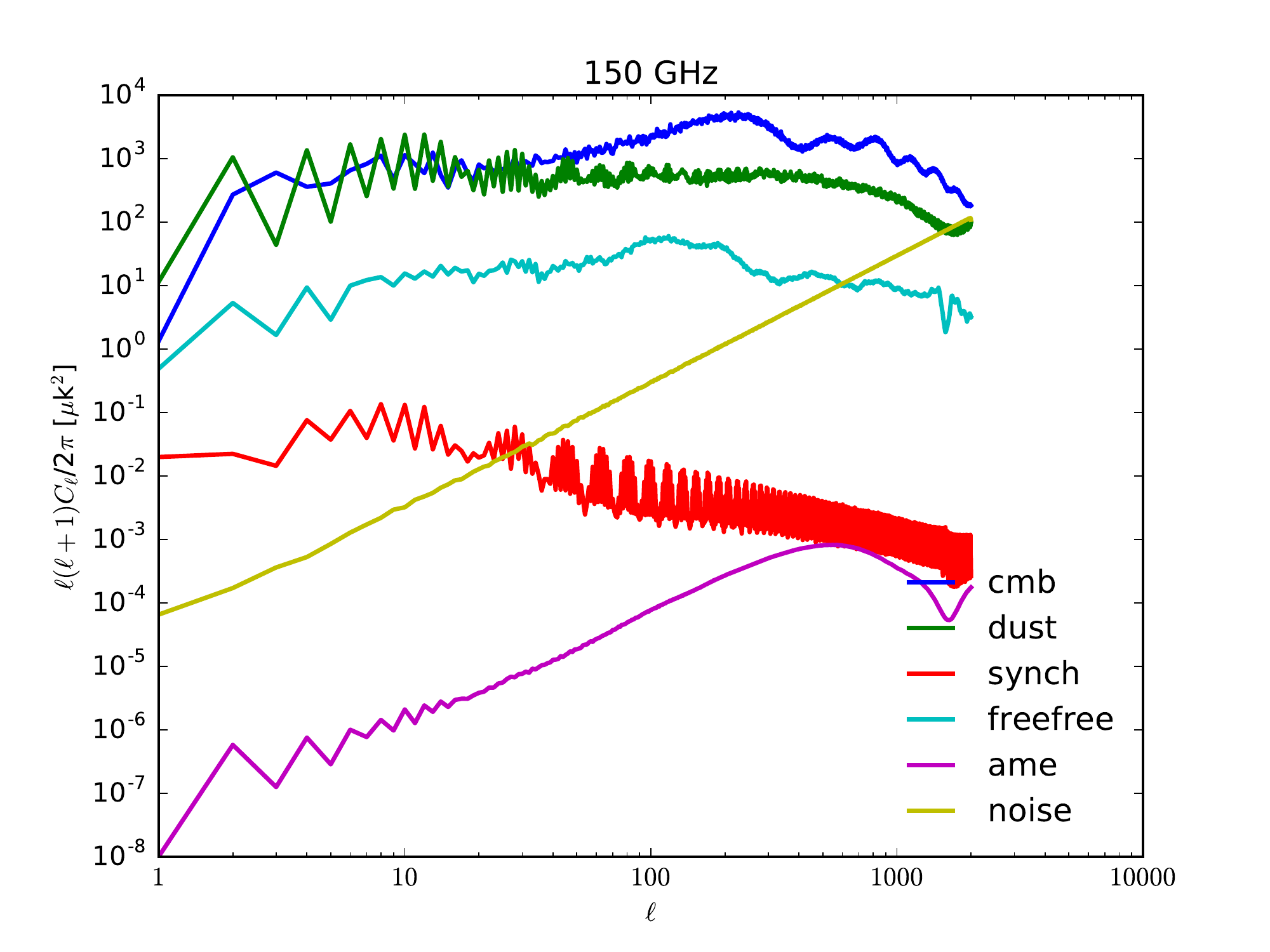}
   }

\subfigure{
   \includegraphics[width=2.3in] {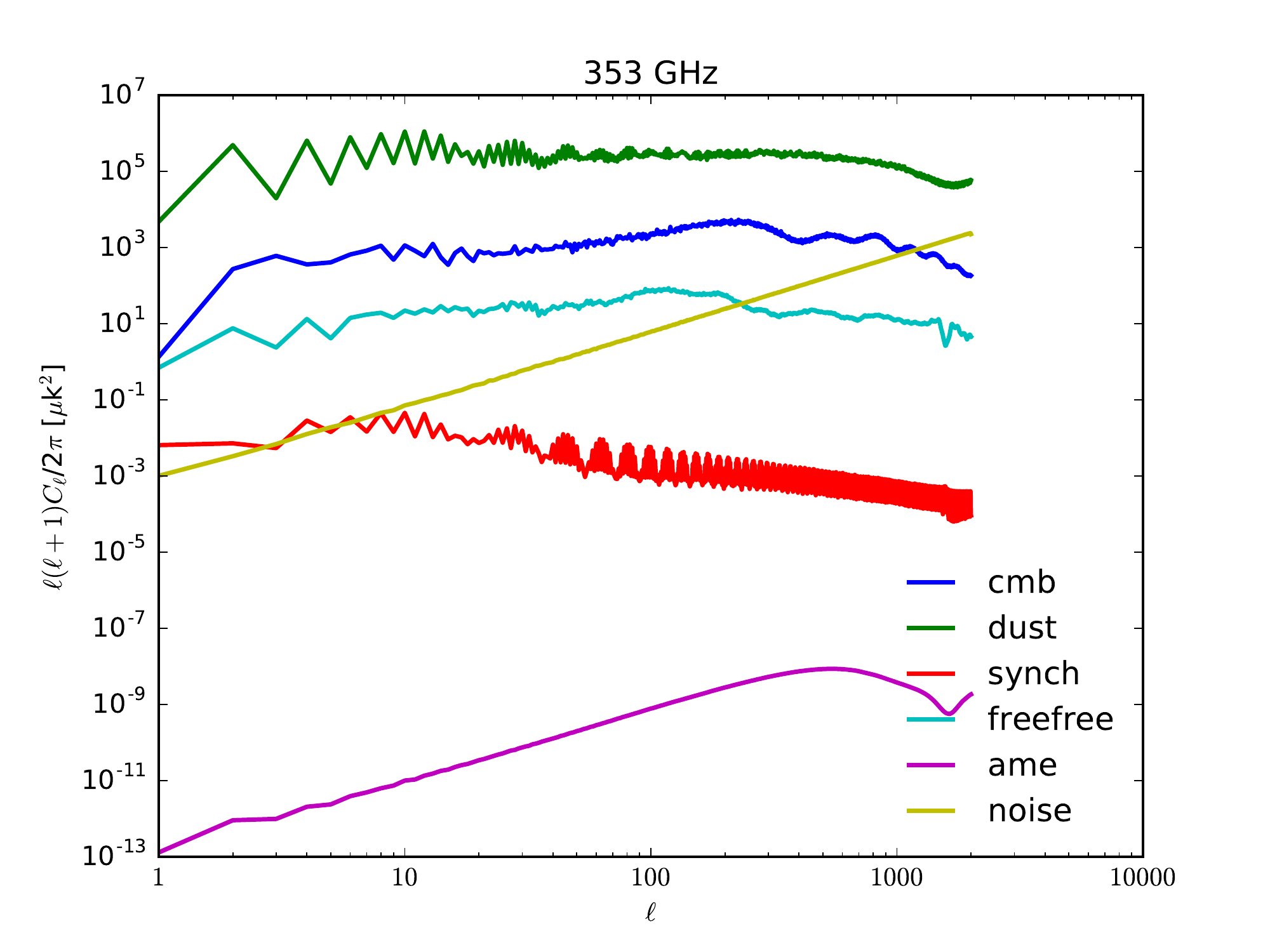}
   }
}

\caption{{\it Upper panels}: Example of realistic simulations of the Planck sky for the case ``Galactic-plane cut'' at the frequencies of 30, 150 and 353 GHz. The Galactic plane in the region of $|b|<10^\circ$ is masked out. Frequency-dependent foreground components correspond to a foreground morphology that varies with frequency. {\it Lower panels}: the corresponding angular power spectra of the simulated sources at the masked Planck maps, in which the CMB, noise, synchrotron, free-free, thermal dust and AME have a different $\ell$-dependence. The units are in $\mu$K.}
\label{fig:map}
\end{figure*}

In order to test the accuracy of the ABS approach, we will apply the estimator in Eq.~\ref{eq:abs1} to realistic simulations. We use full sky Planck simulations provided by~\cite{2017MNRAS.469.2821T}. The maps are generated at the seven frequencies of Planck instruments (30GHz, 44GHz, 70GHz, 95GHz, 150GHz, 217GHz and 353GHz) using HEALPix~\citep{2005ApJ...622..759G} with the resolution of ${N_{\rm side}= 1024}$, for which the spherical harmonics are computed up to the multipole $\ell_{\rm max} =2000$.  In this study, Planck maps in 545GHz and 857GHz are not taken into account, since the inclusion of such two channels can not improve the CMB recovery considerably and thus we can safely neglect them in our test. This is because that the noise levels in these two channels are significant larger than the others and we can not efficiently gain additional useful information of foregrounds from these two channels.

For simplicity, the primary beam pattern in this study is assumed to be unity for all frequency channels. This will not capture realistic observations, whereas all maps can be easily degraded to the same resolution by the beam smoothing process (i.e., a beam convolution). Ideally, if the beam pattern only depends on frequency, this smoothing process, in principle, would not affect the foreground removal. But if the beam depends not only on the frequency but also on the sky direction, such anisotropic beam over the sky might slightly change the source estimations. We will leave this to future works.

Together with the pure CMB signal, four foreground components and Gaussian white noise at the expected level of Planck instruments are mixed in the maps. These state of the art simulations of different Galactic components provide strict tests for the ABS estimator. The CMB signal is randomly generated by using a temperature power spectrum predicted by the public code CAMB~\citep{Lewis2000} with the standard cosmological parameters~\citep{2016A&A...594A..13P}.

Synchrotron, free-free, thermal dust and AME contributions have been taken into account in our simulations and we adopt the nominal \texttt{PySM} model~\citep{2017MNRAS.469.2821T} as the fiducial foreground model. In this model, the synchrotron intensity is a scaling of the degree-scale-smoothed 408 MHz Haslam map~\citep{1981A&A...100..209H,1982A&AS...47....1H,2015MNRAS.451.4311R,2008A&A...490.1093M}, with the spectral index being a direction-dependent power-law. For free-free, the nominal model uses the degree-scale smoothed emission measurement and effective electron temperature  \texttt{Commander} templates by~\cite{2013ApJS..208...20B, 2016A&A...594A..10P}. For thermal dust, the simulations use template maps at 545 GHz in intensity, which are estimated from the Planck data using the \texttt{Commander} code~\citep{2016A&A...594A..10P}. The frequency scaling is modeled as a single component, using the best-fit estimate. For AME, the nominal model is derived from a parametric fit to the Planck data and considers the two-component contributions from the spatially varying and non-varying emissivities~\citep{2009MNRAS.395.1055A,2011MNRAS.411.2750S,2011piim.book.....D}. 

In our simulations, we assume a somewhat idealized noise model that is uncorrelated from pixel to pixel and from channel to channel. For a given $\ell$, the white noise rms levels in the 7 frequency channels from 30 to 353 GHz are derived from~\cite{2008A&A...491..597L}, with $\sigma_{\mathcal{D}}^{\rm noise}(\ell)/\mu {\rm K} = 0.066, 0.065, 0.063, 0.028, 0.015, 0.023, 0.068$, respectively.

Since the level of foreground contamination varies significantly across the whole sky, we apply the ABS to simulated maps with different sky cuts in order to study how robust our results are to the level of foreground contamination. Complementarily, an additional test is also performed by varying the amplitude of the synchrotron emission. In summary, four sets of frequency maps are taken into account:
\begin{enumerate}[(A)]
\item maps with a cut of $|b|<10^\circ$ around the Galactic plane (``{\bf Galactic-plane cut}'', in short);
\item the region within $|b|<10^\circ$ only (``{\bf inside Galactic-plane}''), which has the brightest foreground emission and can be regarded as the worst case;
\item full sky maps without any masks (``{\bf full sky}'');
\item the same as case A, but now with a manually enhanced, by a factor of two, synchrotron emission (``{\bf two-times-stronger synchrotron}'').
\end{enumerate}
The results of the last two cases are given in the Appendix to complement the discussion in the main text.

As an example, Fig.~\ref{fig:map} shows the one realization of our simulated Planck maps for the ``Galactic-plane cut'' case at 30, 150 and 353 GHz, respectively, with masks created by~\cite{2017AJ....153..253M}. The maps at other frequency channels are shown in Fig.~\ref{fig:map-unmask} of the Appendix. For comparison, the power spectra of all simulated sources including the CMB, noise, synchrotron, free-free, thermal dust and AME are shown in the lower panels. As seen, the CMB signal in both frequency and angular-scale dependence behaves significantly different from all other signals.  Furthermore, as the mask we adopted has a sharp cutoff at the boundary, computing the angular power spectrum from partial skies causes spurious oscillations inherent to the spherical harmonic transformation. However, these oscillations can be corrected by standard apodization filtering or taking a large bin size $\Delta \ell$, which have nothing to do with foreground removal.

\section{Tests on simulated skies}\label{sect:test}
In this Section, we show the results obtained by applying the ABS approach to the simulated Planck seven-frequency maps. The estimated CMB power spectrum is obtained by averaging over the results from sky maps with 50 independent realizations of instrumental noise. The associated statistical errors are obtained from its dispersion. The CMB signal and foreground components are fixed in different realizations. The cross power spectrum is binned into bins of width $\Delta l$ = 40 throughout the paper when using the ABS approach.

\begin{figure}[htpb]
\centering
\includegraphics[width=3.6in] {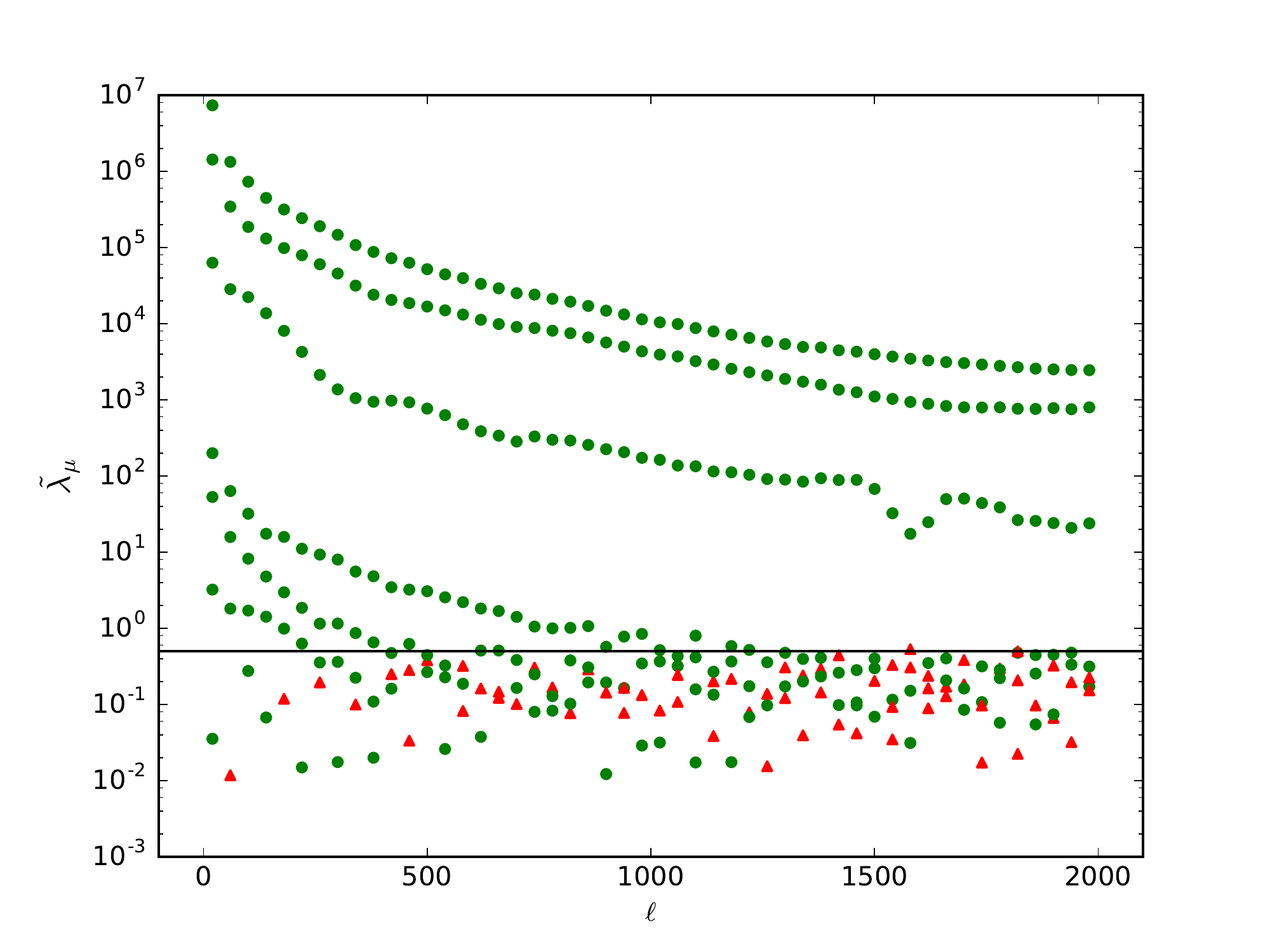}
\caption{Eigenvalues of $\mathcal{\tilde{D}}^{\rm obs}_{ij}(\ell)$ for one realization of our simulated seven-frequency Planck maps with the``Galactic-plane cut'', in which the models of all microwave sources including CMB, 4-component foreground and the noise levels are described in Sect.~\ref{sect:map}. The eigenvalues, $\tilde{\lambda}_\mu$, are shown in absolute value. Due to the instrumental noise, $\tilde{\mathcal{D}}^{\rm obs}_{ij}(\ell)$ is not strictly positive, leading to some small negative eigenvalues (red-triangle). The fiducial threshold $\tilde{\lambda}_{\rm cut} = 1/2$ is shown by the black-solid line. All eigenmodes with the eigenvalues smaller than $1/2$ are excluded from the signal estimation.} 
\label{fig:eigv-mask} 
\end{figure}

\begin{figure}[htpb]
\centering
\includegraphics[width=3.6in] {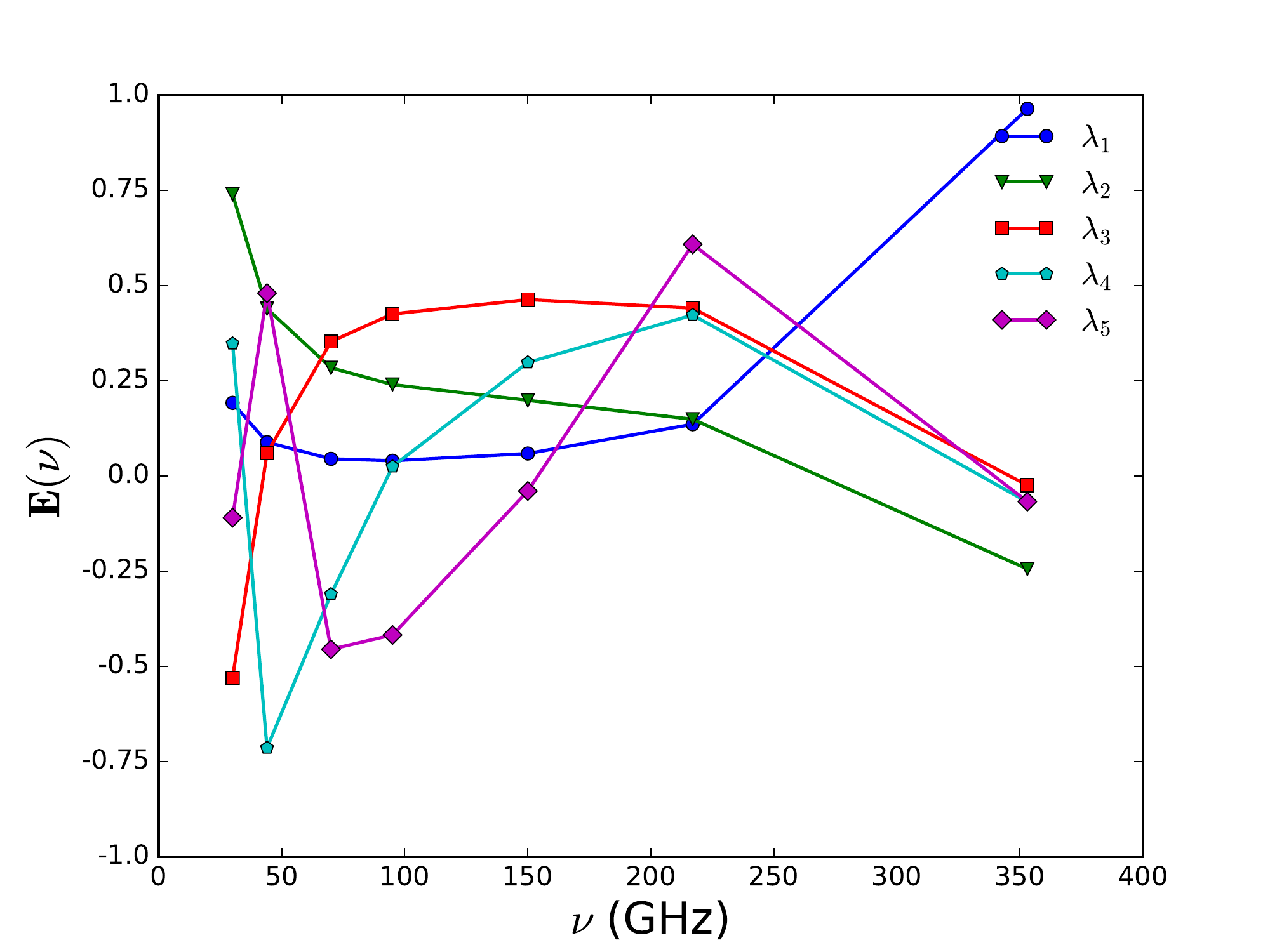}
\caption{The first five eigenvectors ${\bf E}_\mu$ for the maps with Galactic-plane cut as a function of frequency, with the $\ell$ bin centered at $\ell$ = 100. The corresponding eigenvalues are $3.19\times10^5$, $2.03\times10^4$, $4.88\times10^3$, 9.73, 1.12 $\mu$K$^2$, respectively. The last two eigenvectors, which are not shown here, have negligibly small eigenvalues of 0.41 and 0.027 $\mu$K$^2$.  According to Eq.~\ref{eq:abs}, ${\bf E}_\mu$ represents eigenmodes of the raw data set (without normalization by noise variance), and each mode with positive value (i.e. ${\bf E}_{\mu}(\nu)>0$ ) can be qualitatively interpreted as an underlying physical component or a linear combination of such components. The negative ones may have no corresponding physical components because it leads to a reduction in $G_\mu$ defined in Eq.~\ref{eq:abs}. For example, (1) the first eigenvector at $\nu\gtrsim200$ GHz is essentially a contribution from the thermal dust, since its amplitude increases with frequency, which is compatible with the frequency dependence of the thermal dust in our simulation. When $\nu \lesssim200$ GHz, the mode is just a mixture of all of the source signals. (2) The second one at $\nu \lesssim 250$ is possibly dominated by a mixture of the synchrotron, free-free as both of them would monotonically decrease with increasing frequency. (3) The CMB mainly dominates the third one at $70\lesssim \nu\lesssim 200$ GHz as the mode is slowly varying with frequency, which is consistent with the CMB black body spectrum. The remaining eigenvectors show no specific features, with correspondent small eigenvalues receiving contributions from all the sky components.} 
\label{fig:ev-mask-l00} 
\end{figure}

\begin{figure}[htpb!]
\centering
\includegraphics[width=3.6in] {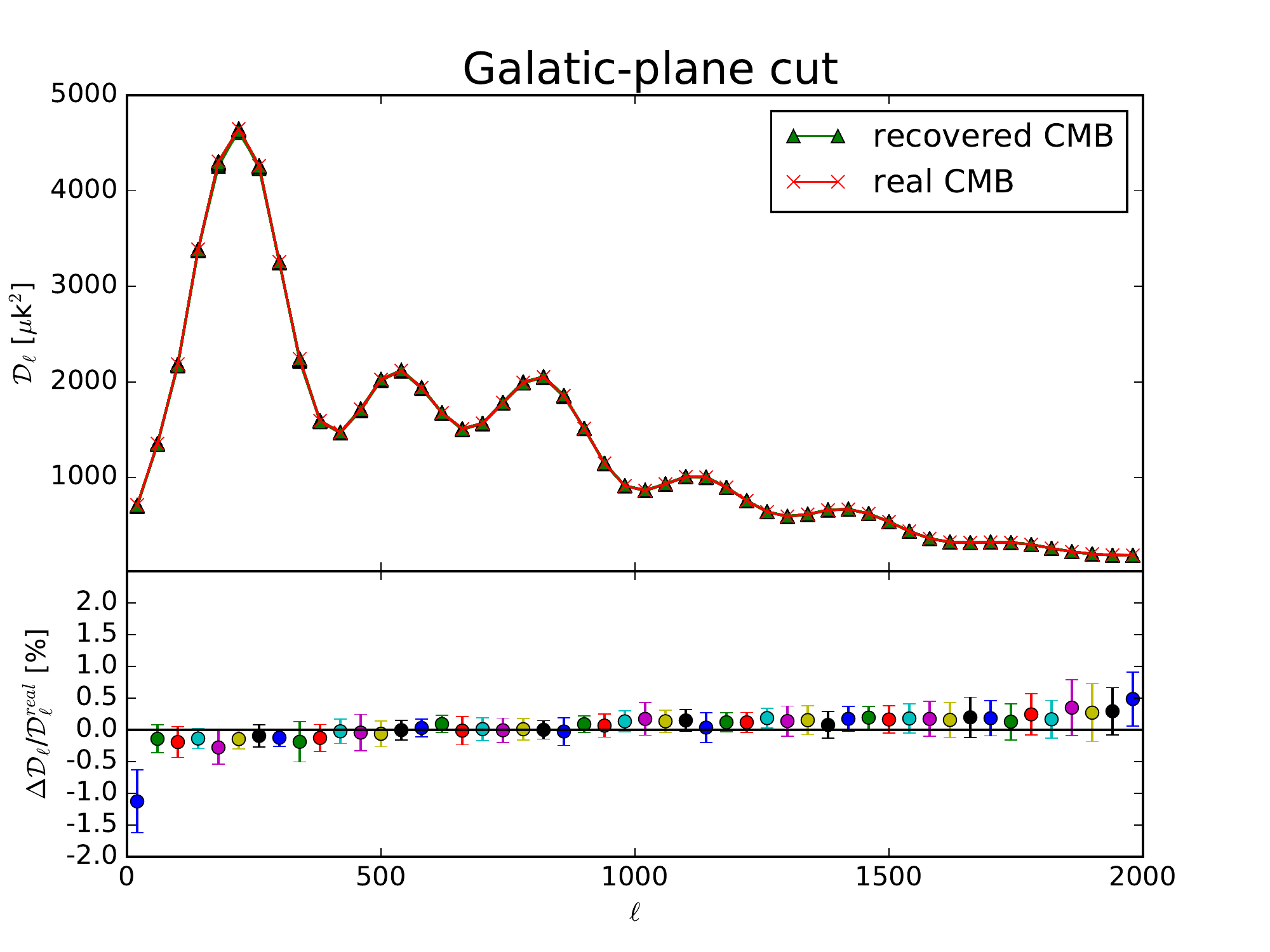}
\caption{The CMB binned power spectrum estimated from the ABS approach from the simulated seven-frequency Planck maps in the case of ``Galactic-plane cut'' where a narrow mask strip around the Galactic plane ($|b|<10^{\circ}$) has been applied to the maps (see Fig.~\ref{fig:map}). The spectrum is binned into bins of width $\Delta l$ = 40. In comparison with the true power spectrum, the relative error, $\hat{\mathcal{D}}^{\rm cmb}_\ell / \mathcal{D}^{\rm real}_\ell -1$, and the associated 1-$\sigma$ statistical error in percentage, are shown in the {\it lower} panel, based on 50 independent realizations of the instrumental noise.}
\label{fig:rec-mask} 
\end{figure}

\subsection{In the case of ``Galactic-plane cut''}
In Fig.~\ref{fig:eigv-mask}, we demonstrate the distribution of the eigenvalues $\tilde{\lambda}_\mu$ and show how $\tilde{\lambda}_\mu$ vary with the multipole $\ell$. For a given $\ell$, there obviously are seven eigenvalues as the order of $\mathcal{\tilde{D}}^{\rm obs}_{ij}$ is $N_f$=7. The complex microwave sky thus can be decomposed into these eigenmodes completely. The first three largest eigenvalues are several orders of magnitude greater than the remaining ones, implying smooth power-law spectral structures in physical foregrounds. Physically, the first three largest eigenvalues are associated with the eigenmodes that are essentially dominated by the free-free, the synchrotron and the total dust (including both thermal dust and AME) emissions, respectively. We also see that these foreground-dominated eigenvalues fall off rapidly as the multipole $\ell$ increase and this fall-off is approximately exponential in the very low-$\ell$ range ($\ell\lesssim80$) where the eigenvalues drop down by an order of magnitude. Since the observed cross band matrix, according to Eq.~\ref{eq:noiseD}, has been normalized by noise, its eigenvalues thus monotonically decrease with increasing $\ell$, which can be estimated from the lower panels of Fig.~\ref{fig:eigv-mask}. Furthermore, the fourth eigenvalues are mainly related with the CMB signal. The fifth eigenvalues might be related to all foreground residuals, together with the CMB component. In addition, the eigenmodes associated with the remaining eigenvalues are essentially dominated by the instrumental noise.

We also see that the last three eigenvalues can reach the noise level, since these eigenmodes are essentially contributions from the noise. Only the noise fluctuations are able to produce negative eigenvalues (red triangles in Fig.~\ref{fig:eigv-mask}) in $\tilde{\mathcal{D}}^{\rm obs}_{ij}$. This is the reason why we choose a threshold $\tilde{\lambda}_{\rm cut}$ to reject the noise-induced eigenmodes and to optimally extract the CMB signal. As expected in~\ref{sect:noise}, the noise-induced eigenvalues have zero mean and a typical value of $|\tilde{\lambda}_\mu|\sim 1$, which is confirmed by our simulations, shown in Fig.~\ref{fig:eigv-mask}.

Since the simualted noise levels are significantly smaller than the CMB signal at low-$\ell$ regime, for simplicity, let us consider an ideal case where the noise is assumed to be absent. According to Eq.~\ref{eq:abs}, the dot product of the vectors, $G_\mu=\tilde{\bf f}\cdot\tilde{\bf E}_\mu$, naturally measures  the projection of the CMB signal vector ${\bf f}$ onto the eigenvector ${\bf E}_\mu$ that is contributed from both foregrounds and CMB. Qualitatively, the dominated physical sources for eigenvectors can be inferred from the frequency dependence and the eigenvalues of ${\bf E}_\mu(\nu)$. In Fig.~\ref{fig:ev-mask-l00}, we show the frequency dependence of eigenvectors ${\bf E}_\mu(\nu)$ at the bin centered at $\ell$= 100 for $\mathcal{D}^{\rm obs}_{ij}(\ell)$. Based on the foreground models in our simulations, the synchrotron and free-free dominate the sky at low frequency ($\nu\lesssim100$ GHz) and the thermal dust at high frequency ($\nu\gtrsim100$ GHz). The former two foreground components decrease as frequency increases and conversely the latter one increases along with frequency. Such property indicates that the first eigenvector at the high frequencies is essentially dominated by the dust emissions, and the second one at the low frequencies are probably contributed from synchrotron and free-free. Note that, when ${\bf E}_\mu(\nu)<0$, it can not be simply interpreted as an underlying physical component or a mixture of all the sky sources because its negative value will lead to a reduction in $G_\mu$ in terms of Eq.~\ref{eq:abs}.

More interestingly, since the third eigenvector at intermediate frequencies ($70\lesssim \nu\lesssim 200$) has a slowly-varying feature over frequency, which is roughly consistent with the emission law of the CMB, we expect it represents the CMB-dominated mode. As a further check, the contribution from the third mode to $\mathcal{D}^{\rm obs}_{ij}$ can then be estimated in terms of the identity, $\mathcal{D}_{ij}^{\rm obs} = \sum_\mu \lambda_\mu {\bf E}_\mu\cdot{\bf E}_\mu^T$, which is about $\lambda_3 {\bf E}_3\cdot{\bf E}_3^T\sim 1.2\times10^3~\mu\rm{K}^2$ where $\lambda_3 = 4.88\times10^3~\mu\rm{K}^2$ and each element of ${\bf E}_3\sim0.5$ from Fig.~\ref{fig:ev-mask-l00}. This value is well compatible with the amplitude of the CMB band power at $\ell\sim100$, $\mathcal{D}^{\rm cmb}_{ij}\simeq 10^3~\mu\rm{K}^2$ (see Fig.~\ref{fig:map}). In addition, there are no specific behaviors for the other eigenmodes, suggesting that they are dominated by a combination of all physical sources.   

In Fig.~\ref{fig:rec-mask}, we present the recovered CMB band powers for the case ``Galactic-plane cut''. The estimated band powers are obtained by averaging over the results from 50 simulated seven-frequency maps with independent realizations of the instrumental noise,  and the associated statistical error in each $\ell$-bin (i.e., standard deviation) is computed by the total dispersion of all realizations divided by the square root of the number of realizations. As seen, the ABS estimator provides a precise foreground subtraction. The comparison of the recovered spectrum with the true one shows a perfect agreement over the whole multipole range.

In order to quantitatively appreciate the differences between the recovered and the true one, we show in Fig.~\ref{fig:rec-mask} at the lower panel the relative error, $r$, in percentage, where $r= \hat{\mathcal{D}}^{\rm cmb}_\ell / \mathcal{D}^{\rm real}_\ell -1$ and $\hat{\mathcal{D}}^{\rm cmb}_\ell$ is the estimator of the CMB band powers in Eq.~\ref{eq:abs1}. In both large and small scales, the ABS estimator provides an extremely good recovery of the CMB band power spectrum. The ensemble-averaged relative error for each $\ell$ bin, $\left<r\right>$, is negligibly small. The averaged relative error over all 50 $\ell$-bins is 0.047\%, which is much smaller than the averaged 1-$\sigma$ statistical uncertainty of $\sigma_r=0.0023$. At high-$\ell$ bins ($\ell\gtrsim$ 1000) where the noise level is slightly above or comparable with the CMB signal, the error bars become larger than those at the low $\ell$-bins and increase as $\ell$ increase. Due to the noise effect, our estimator shows a small positive bias toward the higher $\ell$ region. The maximum relative error of $\left<r\right> = -0.011$ is at the lowest $\ell$-bin (2$\leq \ell<40$), which is not surprising as the maximum foreground contamination relative to the CMB occurs at very large scales. Furthermore, the corresponding absolute errors with respect to the true band powers is given by $\left<\Delta\mathcal{D}\right>$ with $\Delta\mathcal{D}= \hat{\mathcal{D}}^{\rm cmb}_\ell - \mathcal{D}^{\rm real}_\ell$. On average over all $\ell$-bins, the absolute error is about $-0.56~\mu\rm{K}^2$ with 1-$\sigma$ error of 2.66 $\mu\rm{K}^2$, so that the ABS estimator provides an accuracy at the level of 1 $\mu\rm{K}^2$ on the power spectrum recovery.

In Tab.~\ref{tab:cl}, the relative and absolute errors and associated 1-$\sigma$ uncertainties for six example bins show the typical accuracy level. The above results indicate that the estimated amplitudes of most $\ell$ bins (except for the first and the last ones) are consistent with the input values within 1-$\sigma$, strongly implying that the ABS approach provides an unbiased estimate of the CMB spectrum over almost all scales. 


It is worth thinking about the reasons for the negative and positive biases in the recovered band powers. One can see that the negative bias appears to be strongest at the lowest $\ell$ bin, and the biases become positive at small scales and increase in magnitude with increasing $\ell$. From the simulations, the measurements suffer the strongest foreground contamination with respect to the signal at the lowest $\ell$ bin. One would naively expect that the residual foreground contamination would cause a positive bias in measurements. This is true only for the case that the number of frequency channels is extremely insufficient, such as assuming 2 channels of 30 GHz and 353 GHz in the survey. Referring to Eq.~\ref{eq:abs1}, the contribution of the $\mu$-th eigenmode to $\mathcal{\hat{D}}^{\rm cmb}$ is inversely proportional its eigenvalue $\tilde{\lambda}_\mu$, so that eigenmodes with small eigenvalues would significantly affect the signal estimation. In this two-channel survey, the eigenvalues ($\tilde{\lambda}_1\sim10^7~\mu\rm{K}^2$ and $\tilde{\lambda}_2\sim10^5~\mu\rm{K}^2$) are significant greater than the value of the cut off we used, and thus the CMB signal is essentially determined by the second eigenmode, i.e.,  $\mathcal{\hat{D}}^{\rm cmb}\approx \tilde{\lambda}_2/\tilde{G}^2_\mu\sim \tilde{\lambda}_2$ in terms of Eq.~\ref{eq:abs1}, causing a significant overestimate on the CMB.  

On the other hand, the seven-frequency survey has provided the necessary frequency coverage essentially, the negative deviation may be induced by a potential correlation arising between the CMB signal and foregrounds at large scale. Because of such correlation, the signal would be poorly orthogonal to the foreground in frequency space, leading to large values of $\tilde{G}_{\mu}$ and then to a small value of $\mathcal{\hat{D}}^{\rm cmb}$. We further investigate the changes of the negative bias by varying $\tilde{\lambda}_{\rm cut}$. The simulations show that, all the recovered signals are underestimated at the first $\ell$ bin, with the relative deviations of about $-4.97\%, -1.1\%, -0.59\%$ and $-0.53\%$ for $\tilde{\lambda}_{\rm cut}=$ 0, 0.5, 10 and 100, respectively. The existence of the negative bias can be therefore qualitatively understood from these results.


The positive biases in the high-$\ell$ region are mainly due to the fact that the noise will gradually dominate over the CMB signal when increasing $\ell$, and noise-dominated unphysical eigenmodes with eigenvalues $\tilde{\lambda}$ below $\tilde{\lambda}_{\rm cut}= 1/2$ are excluded in the determination of the CMB. As known, all eigenmodes with positive eigenvalues always have positive contributions to the summation of terms $\tilde{G}^2_{\mu}/\tilde{\lambda}_{\mu}$. Due to a  positive cut-off ($\tilde{\lambda}_{\rm cut}>0$), some positive contributions are not included in computing the sum, directly resulting in a decrease in the sum and then an overestimate in the measurement of the CMB. As seen from Figs.~\ref{fig:eigv-mask} and~\ref{fig:rec-mask}, the last four eigenmodes are completely excluded at $\ell\gtrsim1300$, and a positive biases essentially arise at the same scales, which is well consistent with our explanation.


\begin{table}
\begin{center}
\begin{tabular}{c|c|c|c|c}
\hline
Range of multipoles    &$\left<r\right>$&  $\sigma_r$ & $\left<\Delta\mathcal{D}\right>$ ($\mu \rm{K}^2$)  & $\sigma_{\Delta \mathcal{D}}$  ($\mu \rm{K}^2$) \\
\hline
2 - 40 &  -0.011  & 0.0049 &  -8.03 & 3.51 \\
160 - 200   &  -0.0026  & 0.0027 &  -11.2 & 11.9 \\
480 - 520   &  -0.0006  & 0.0021 &  -1.24 & 4.11 \\
960 - 1000   &  0.0013  & 0.0016 &  1.22 & 1.52 \\
1480 - 1520   &  0.0016  & 0.0021 &  0.88 & 1.15 \\
1960 - 2000  &  0.0049  & 0.0042 &  0.91 & 0.78 \\
\hline
\end{tabular}
\end{center}
\caption{Relative and absolute errors on the recovered CMB band power spectrum for the ``Galactic-plane cut'' maps at several $\ell$ bins examples.}
\label{tab:cl}
\end{table}

\begin{figure*}[htpb]
\centering
\mbox{
 \subfigure{
   \includegraphics[width=2.3in] {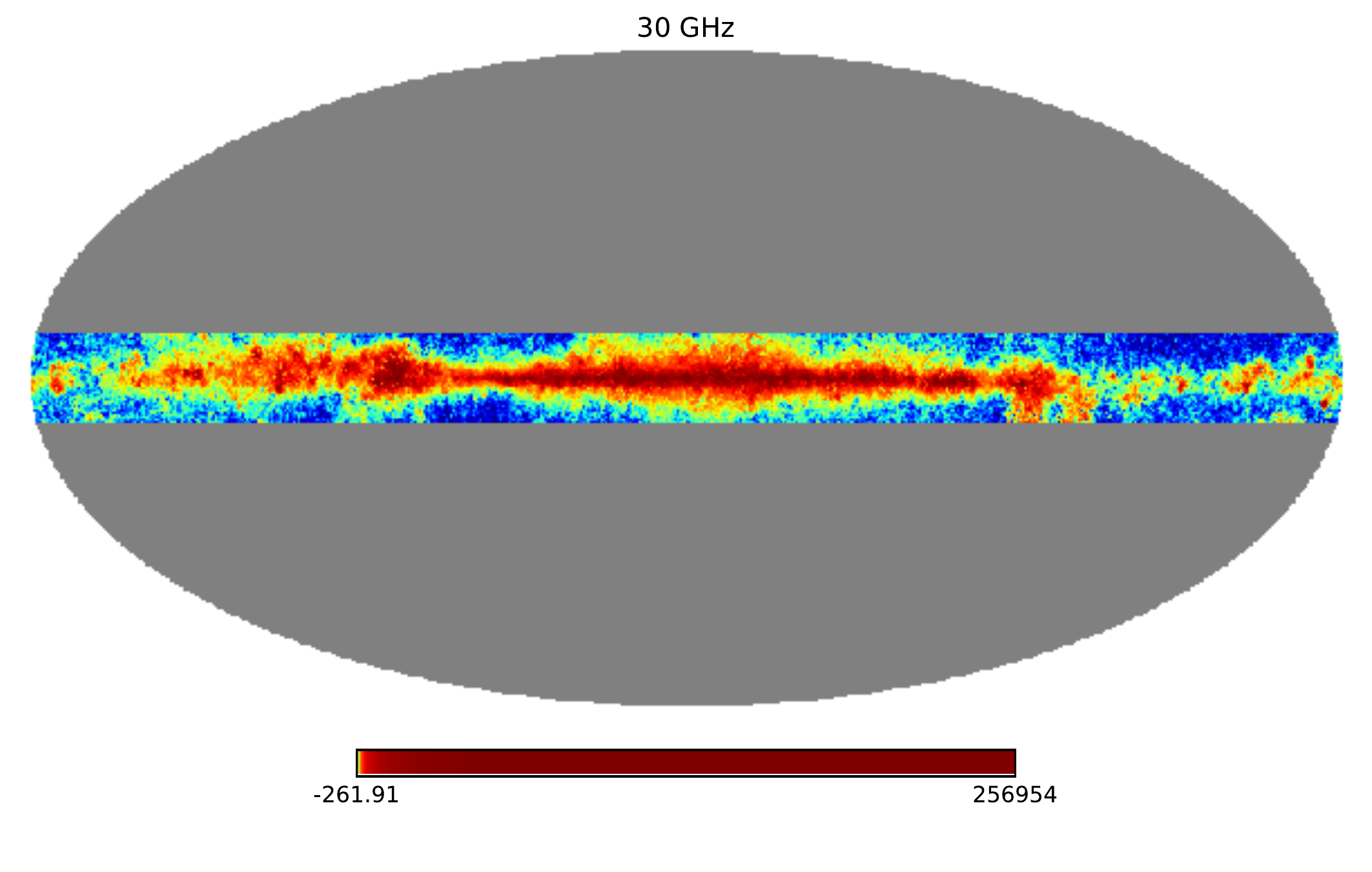}
   }

\subfigure{
   \includegraphics[width=2.3in] {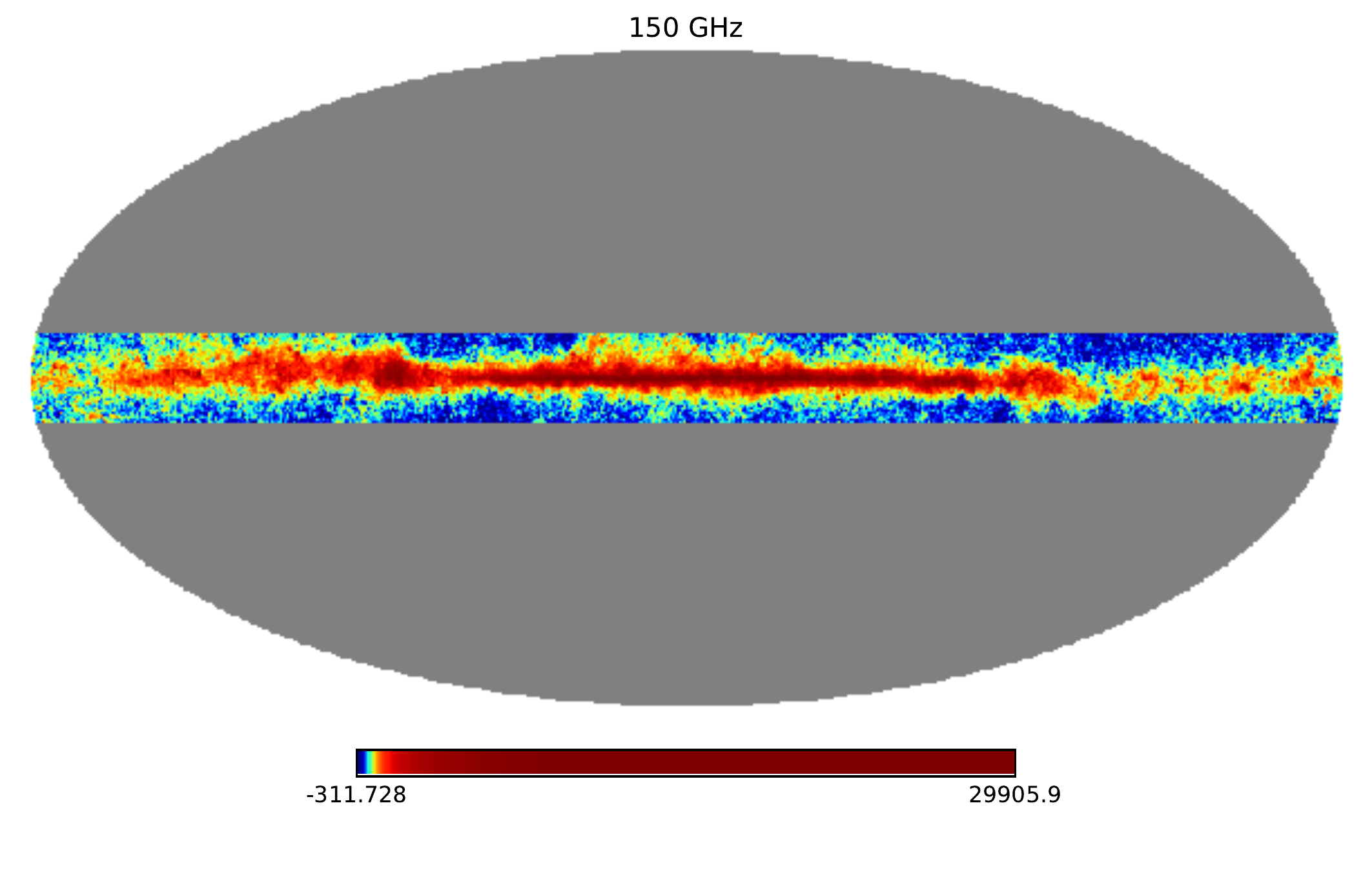}
   }

\subfigure{
   \includegraphics[width=2.3in] {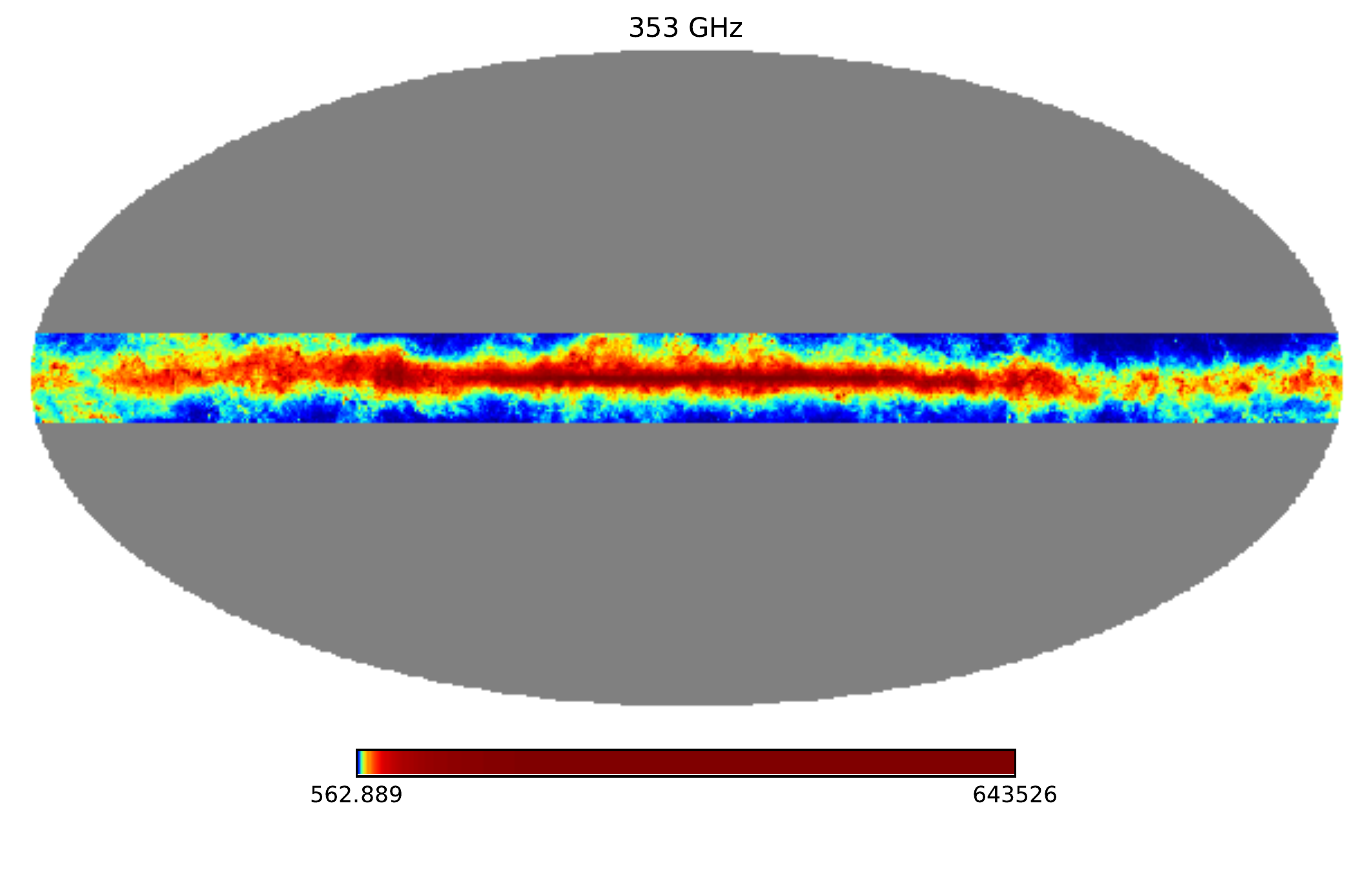}
   }
}

\mbox{
 \subfigure{
   \includegraphics[width=2.3in] {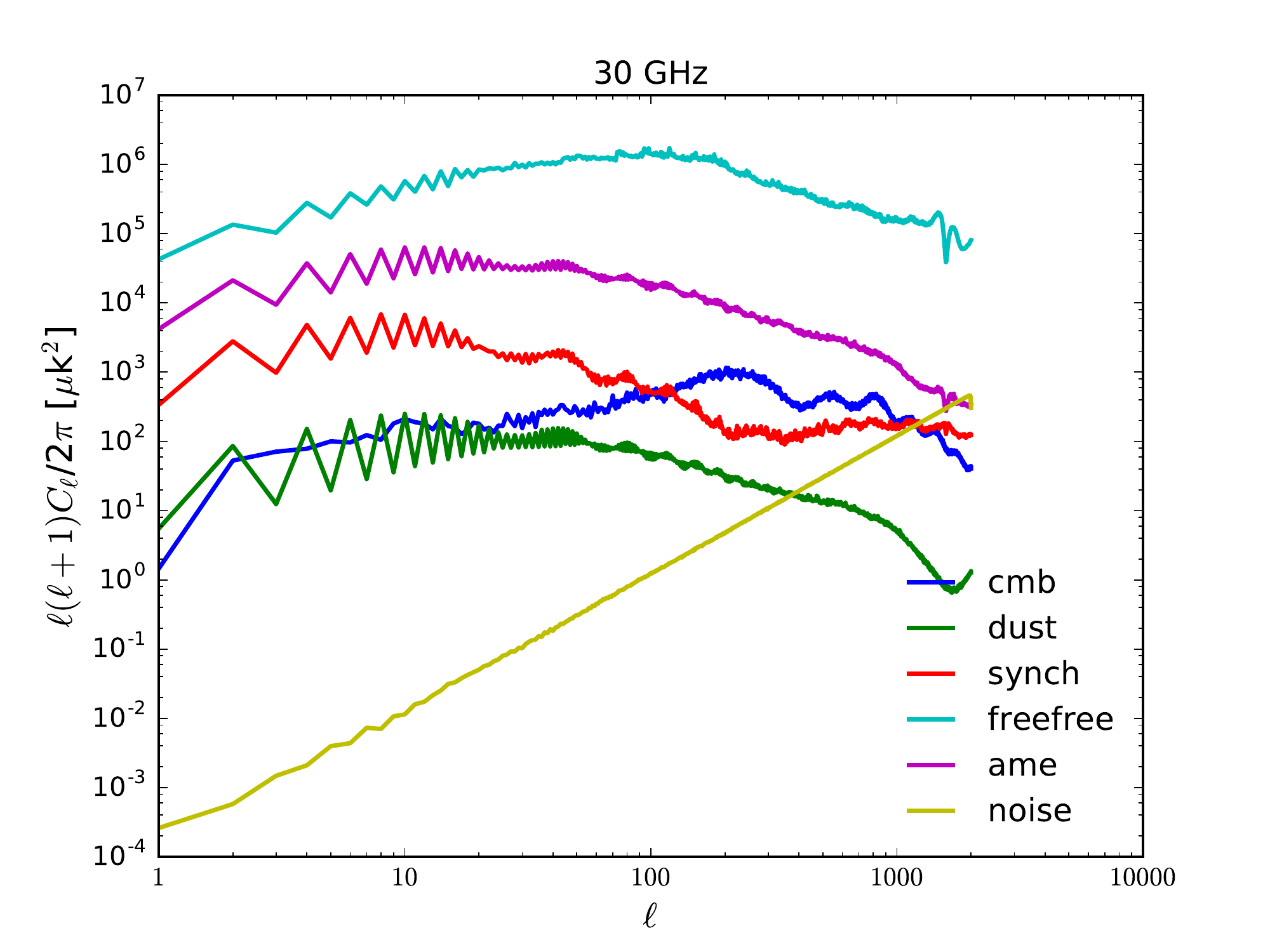}
   }

\subfigure{
   \includegraphics[width=2.3in] {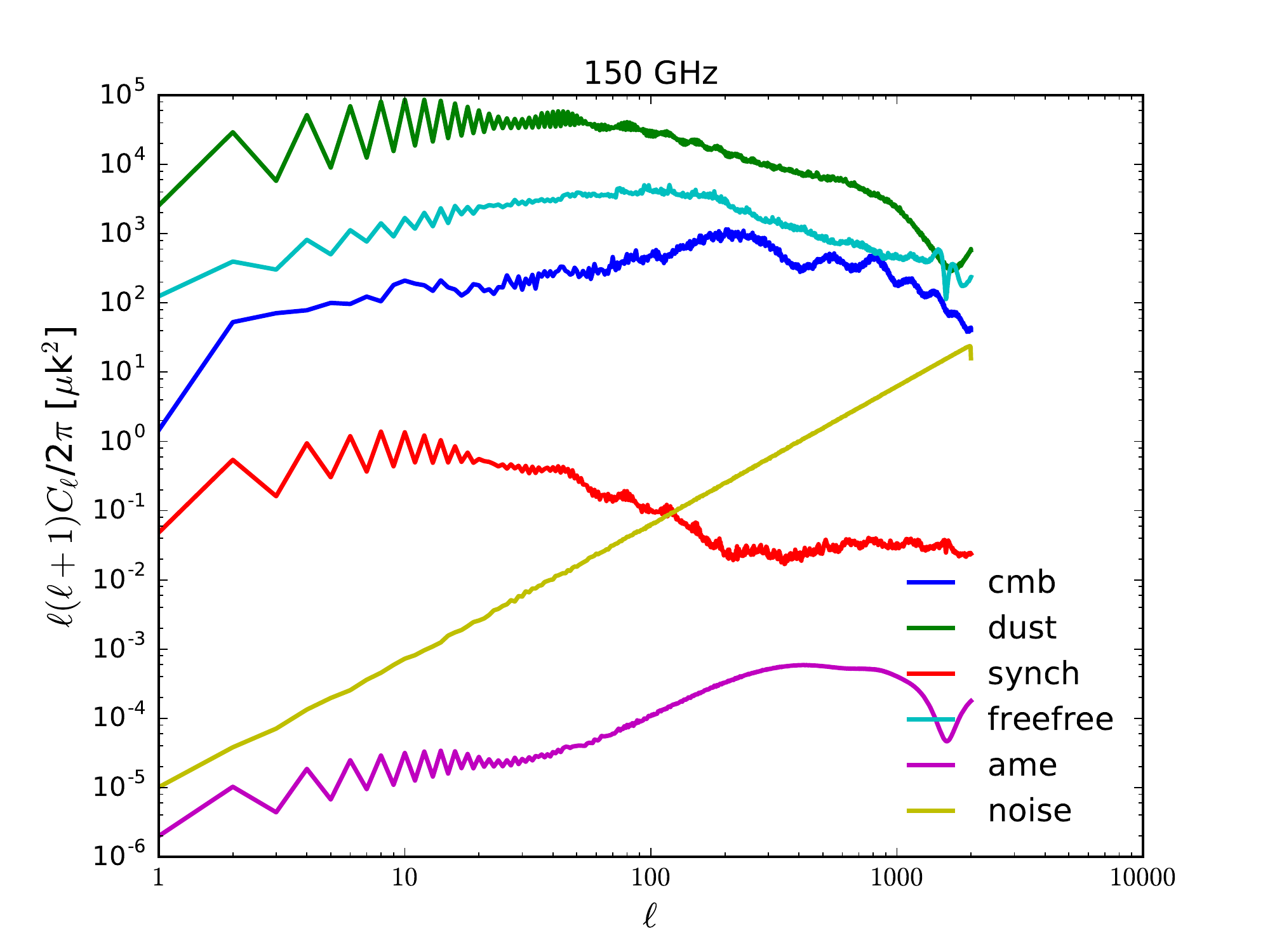}
   }

\subfigure{
   \includegraphics[width=2.3in] {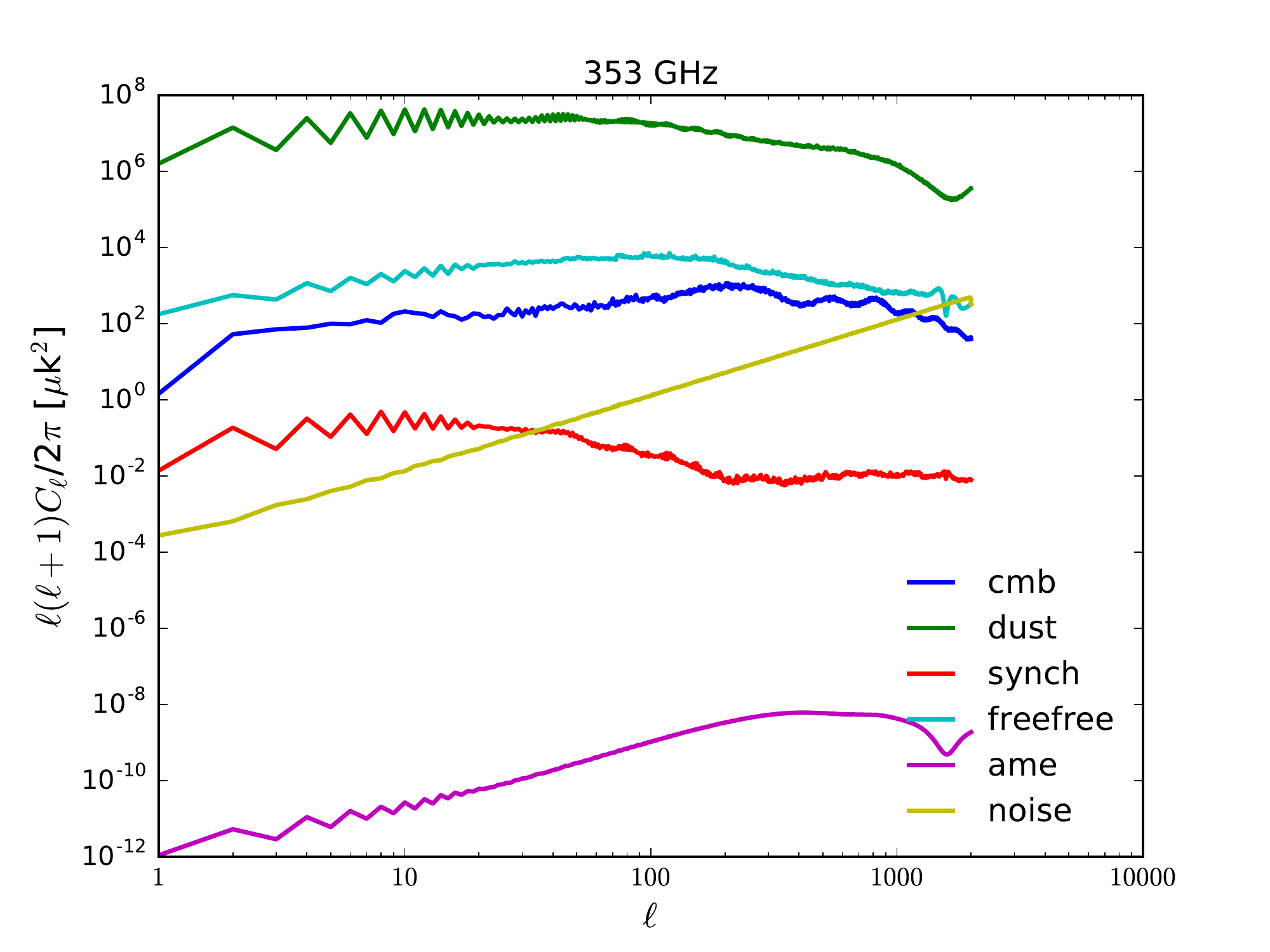}
   }
}

\caption{Same as Fig.~\ref{fig:map}, but for the case ``inside Galactic-plane'', where only the region with $|b| < 10^{\circ}$ is used for CMB band power spectrum estimation.}
\label{fig:map-out}
\end{figure*}

\begin{figure}[htbp]
\centering
\includegraphics[width=3.6in]{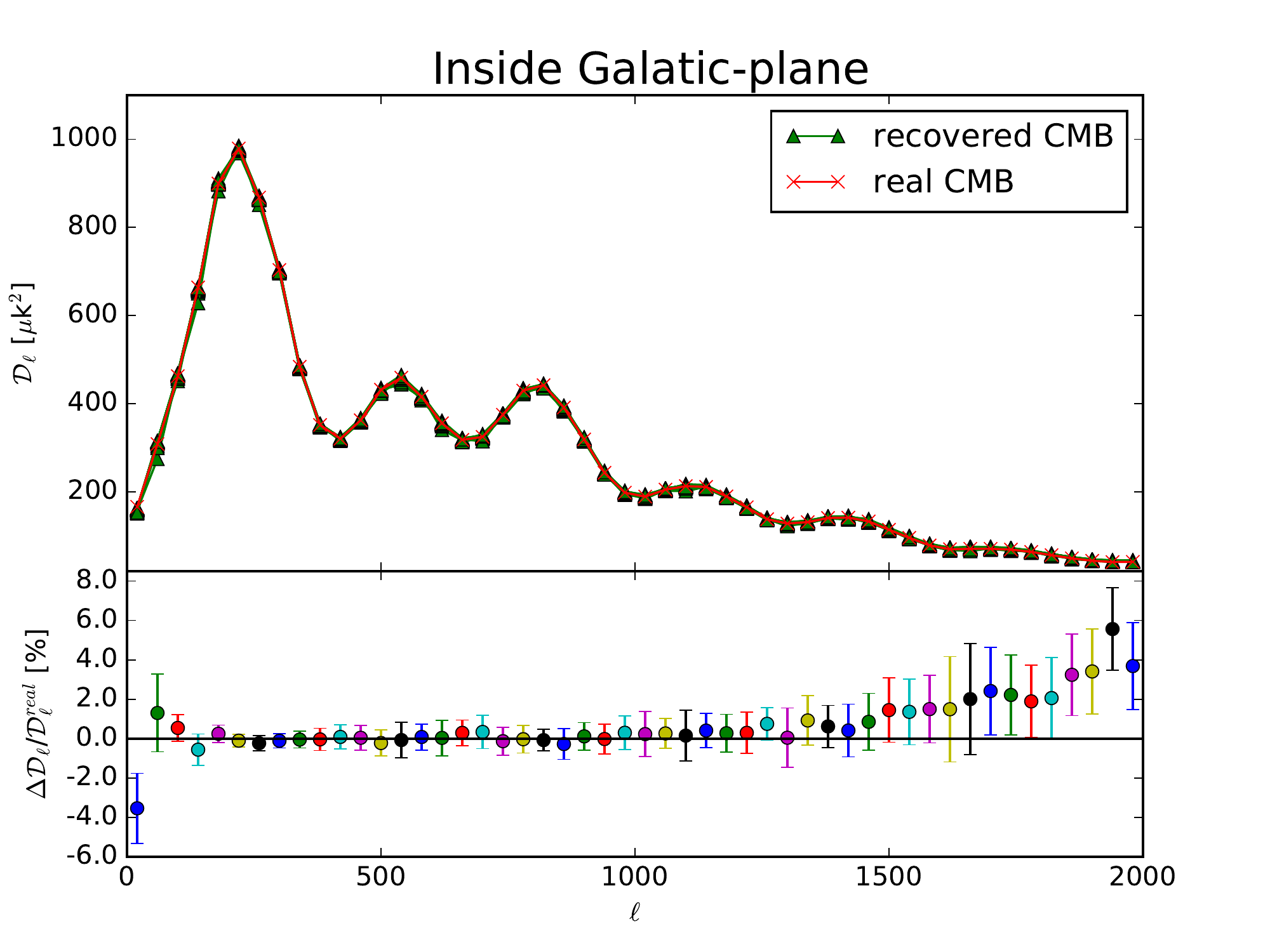}\label{fig:maskout}
\caption{Same as Fig.~\ref{fig:rec-mask}, but for the case ``inside Galactic-plane'' (shown in Fig.~\ref{fig:map-out}), which is an extreme case suffering the brightest foreground emissions.}
\label{fig:rec-out}
\end{figure}

\subsection{In the case of ``inside Galactic-plane''}
To demonstrate the validity of the ABS approach, we test it against the most extreme case (``inside Galactic-plane'') for which only the Galactic plane is used for the CMB estimation and all the regions outside of  $|b|<10^{\circ}$ are masked out (see Fig.~\ref{fig:map-out}). As expected, the CMB signal is highly contaminated by the brightest foreground emissions within the Galactic plane, which would much worsen the accuracy of the recovery.

As seen in Fig~\ref{fig:rec-out}, the mean relative errors $\left<r\right>$  become relatively large for low- and high-$\ell$ bins, which reach about $-3.5\%$ and $5.7\%$ at the bins centered at $\ell$= 20 and 1940, respectively. The reason relies on the fact that these large and small scales correspond to the foreground- and the noise-dominated regions, respectively, where foreground and noise may leak into the estimated signal. In addition, the true power spectrum of the CMB inside the Galactic-plane is much smaller than that in the case ``Galactic-plane cut'' (see Fig.~\ref{fig:map-out}), which would enlarge the relative error even if the absolute error is the same as that in the case ``Galactic-plane cut''. We also find the recovered spectrum in the multipole range of $100\lesssim\ell\lesssim1400$ is almost the same as the true one with only $<1\%$ deviation (e.g., $\sim0.2\%$ at $\l\sim180$ and $\sim0.03\%$ at $\l\sim480$; see Tab.~\ref{tab:cl-inside}), which are comparable with those of the case ``Galactic-plane cut''.

At $\ell \gtrsim 1500$, the relative errors and associated error bars are somewhat increased, since the noise signal dominates in this high-$\ell$ region, slightly overestimating the CMB signal. By averaging over all bins, we find the relative error of $\left<r\right> = 0.76\%$ and associate statistical uncertainty of $\sigma_r = 0.011$, corresponding to the absolute error of 0.51 $\mu{\rm K}^2$ with the uncertainty of 2.16 $\mu{\rm K}^2$.
Even for this extreme case, the ABS approach can still provide the estimate at below $1\%$ level. The averages of the relative errors and statistical uncertainties become 16 and 5 times larger than those of the case ``Galactic-plane cut''. The average of the absolute errors however is only slightly increased by $10\%$. It implies that the ABS estimator is capable of identifying and removing all the foreground sources from the simulated Planck seven-frequency maps at the level of about 0.5 $\mu{\rm K}^2$ by averaging over all bins. The ABS estimator is extremely robust and different sky cuts cannot bias the estimate significantly.

\subsection{The Null Test}\label{sect:null}

\begin{figure}[htpb]
\centering
\includegraphics[width=3.6in]{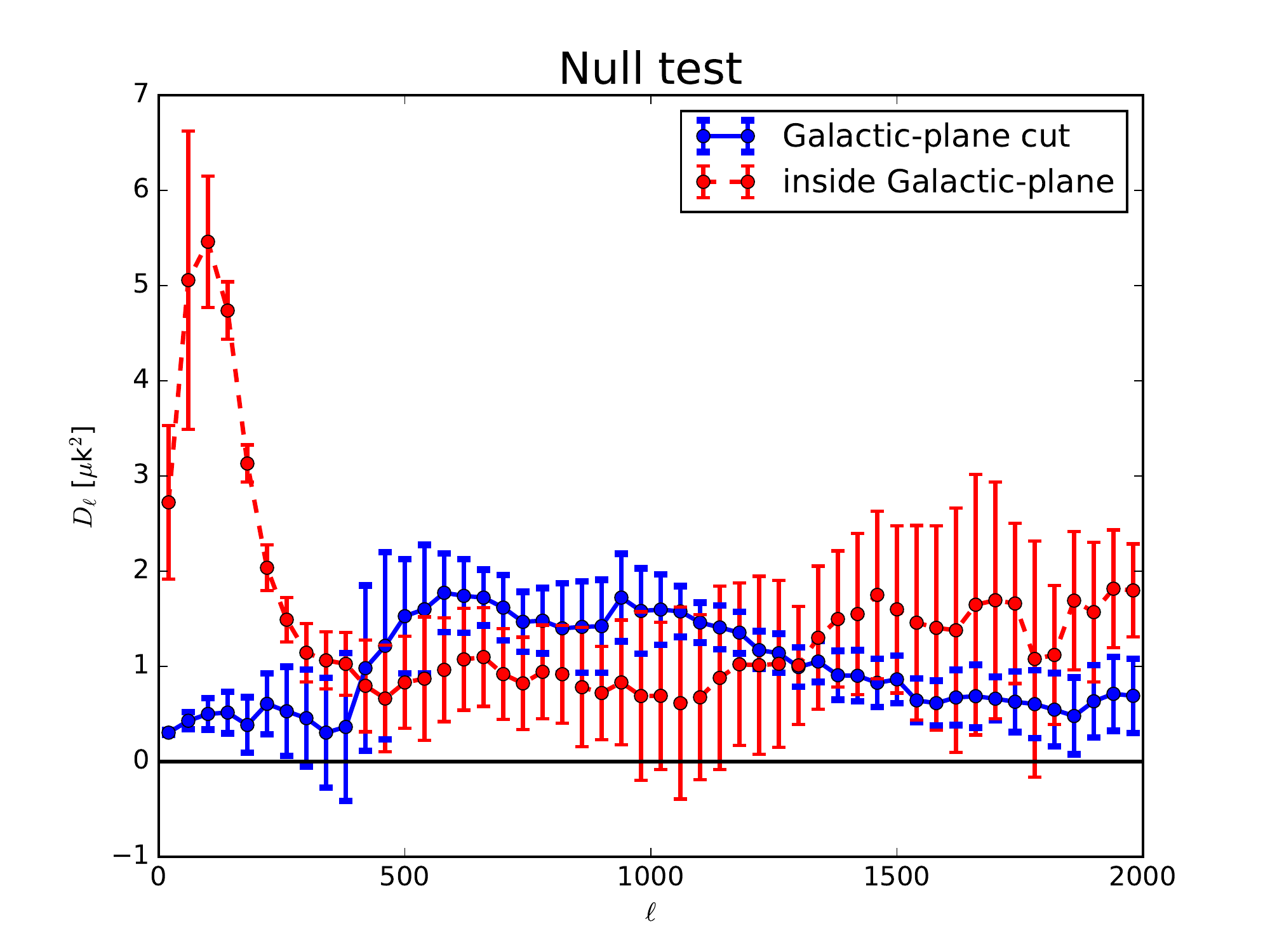}
\caption{The result for the null test with the ABS method in the cases of ``Galactic-plane cut'' and ``inside Galactic-plane''. The ABS-derived power spectrum and the associated 1-$\sigma$ statistical error are estimated from 50 independent realizations of the instrumental noise. The shift parameter is set to be large enough (e.g., $\mathcal{S}\sim 50 \sigma_{\mathcal{D}}^{\rm noise}$) to stabilize the calculation, yielding in completely convergent results.}
\label{fig:null}
\end{figure}


The null test is expected to be an important check for verifying the validity of our estimator, especially for detecting the extremely faint primordial B-mode polarization. To investigate our null test, we apply the ABS-estimator to the simulated sky maps in the absence of the CMB signal. The results of the null test in the ``Galactic-plane cut'' and ``inside Galactic-plane'' cases are shown in Fig.~\ref{fig:null}.  We see that there are no significant false signals for both cases, and the astrophysical sources and noise can bias the estimate at the level of $\sim 1~\mu\rm{K}^2$ on average, which is comparable with the bias level estimated from the maps in the presence of the CMB signal.

 In order to accurately evaluate the null test, for each $\ell$-bin we calculate the statistic, $\chi^{\rm null}_{\ell} \equiv \mathcal{D}^{\rm null}_{\ell}/\sigma_\ell$, where $\mathcal{D}^{\rm null}_{\ell}$ is the null band power and $\sigma_\ell$ is the standard deviation of $\mathcal{D}^{\rm null}_{\ell}$ in the 50 realization of our simulations. We evaluate $\chi^{\rm null}_{\ell}$, which is responsive to systematic biases in the null spectra and provides a quantitatively analysis to verify the null test. The mean of the $\chi^{\rm null}_{\ell}$ over all bins is $3.37\pm 2.04$ in the case of ``Galactic-plane cut'', and is $2.69\pm 3.11$ in ``inside Galactic-plane'', which validates that the null-power-spectra are consistent with zero in 1.7-$\sigma$ and within 1-$\sigma$, respectively. We also find the mean of the $\chi^{\rm null}_{\ell}$ for the ``full sky'' case is $5.86\pm 8.1$, consistent with zero in 1-$\sigma$. To further check the consistency and stability, we perform the null tests for the maps with increasing the mask. There yield $\chi^{\rm null} = 2.25\pm 3.25$ for the Galactic mask of $|b| < 20^{\circ}$, and $0.88\pm 1.54$ for $|b| < 40^{\circ}$. Thus, masking out larger Galactic-plane region would lead to smaller systematic bias, and all the null spectra for the above five null tests are below 2-$\sigma$. We can therefore conclude that the ABS estimator successfully passed all null tests performed here and  demonstrates a stable performance against the skies with different masks.

As seen, some large false amplitudes occur at $\ell\lesssim 400$ in the case ``inside Galactic-plane'' in which the maximum one reaches 5.3 $\mu\rm{K}^2$ at $\ell\simeq 100$, about 5-$\sigma$ far away from the true value. As known, the foreground contamination at the Galactic central region is strongest while having much stronger correlation at large scale than at small scale, so that such contaminants may slightly bias the signal estimate at low-$\ell$ region. Thanks to the robustness and effectiveness of the ABS, the recovered amplitudes are still consistent with zero within a few $\mu\rm{K}^2$, even for this worst case.  Nonetheless, in practice, one has to exclude this highly contaminated region to avoid any possible biases induced from foregrounds.         


\begin{table}
\begin{center}
\begin{tabular}{c|c|c|c|c}
\hline
Range of multipoles    &$\left<r\right>$&  $\sigma_r$ & $\left<\Delta\mathcal{D}\right>$ ($\mu \rm{K}^2$)  & $\sigma_{\Delta \mathcal{D}}$  ($\mu \rm{K}^2$) \\
\hline
2 - 40 &  -0.035  & 0.017 &  -5.88 & 2.95 \\
160 - 200   &  0.0025  & 0.0044 &  2.22 & 3.95 \\
480 - 520   &  -0.00037  & 0.0056 &  -0.13 & 1.98 \\
960 - 1000   &  0.0085  & 0.0031 &  0.61 & 1.69 \\
1480 - 1520   &  0.015  & 0.016 &  1.71 & 1.83 \\
1960 - 2000  &  0.039  & 0.021 &  1.6 & 0.87 \\
\hline
\end{tabular}
\end{center}
\caption{Same as Tab.~\ref{tab:cl}, but for the case ``inside Galactic-plane''.}
\label{tab:cl-inside}
\end{table}

\section{Conclusions}\label{sect:con}
We have tested the ABS estimator for recovering the CMB power spectrum. Our estimator provides a blind way to analytically extract the CMB power spectrum from foreground contaminated maps by using the measured cross band power between different frequency channels. This estimator does not rely on any assumptions of foreground components and it is only based upon the fact that the CMB follows a blackbody spectrum.

The ABS estimator was applied to simulated muti-frequency Planck maps at 30, 44, 70, 95, 150, 217 and 353 GHz. We keep the simulated foreground as realistic as possible. The microwave sources in the simulations include the CMB, Galactic synchrotron, free-free, thermal dust and  anomalous microwave emission together with instrumental noise. The various components have significantly different angular morphology, frequency dependence and amplitudes.

The results are quite promising. By performing 50 independent realizations of a set of simulations and comparing with the true simulated map, we find that the ABS estimator provide an unbiased and efficient estimate of underlying the CMB power spectrum well within 1-$\sigma$ error bar at most scales. When choosing a Galactic mask excluding the region $|b| < 10^{\circ}$, the CMB power spectrum is recovered with an accuracy at the level of less than 0.5\% over all scales. Moreover, we have tested the ABS estimator against more extreme situations: (1) full sky maps without any mask, (2) adding synchrotron foreground maps increased by a factor 2, and (3) inside a narrow strip of sky around the Galactic plane ($|b|<10^{\circ}$). We find that the performance of the ABS estimator is remarkably robust. It can recover the CMB power spectrum accurately in the presence of foregrounds, reaching relative errors of order a few percent.

It is interesting to evaluate the recovery performance by comparing our results with the Planck 2015 temperature power spectrum~\citep{2016A&A...594A..13P}. The residuals with respect to the best-fit theoretical prediction and associated 1-$\sigma$ uncertainties are about tens to hundreds of $\mu\rm{K}^2$ at multipoles $\ell\lesssim500$, and they decrease to several to tens of $\mu\rm{K}^2$ in the range of $500\lesssim\ell\lesssim2500$. Both residuals and uncertainties in Planck are much greater than our results ($\sim1~\mu\rm{K}^2$). Foreground removal with ABS seems to be more robust and effective against existing methods used in Planck. However, evaluating the recovery performance for ABS against the Planck results is complicated and difficult, because (1) the CMB signal is always fixed in our simulations and the cosmic variance induced errors are not taken into account; (2) the primary beam is assumed to be unity for all simulated maps, with no boost in noises at high-$\ell$ regime from beam deconvolution; (3) the simulations cannot model the foregrounds and noise properties as well as systematic effects of the real Planck data to sufficiently high accuracy. It is thus important to further test the ABS approach by using the real Planck data. We will leave this to future works.

We have to mention that the simulated observations in this study are somewhat idealized, without primary beam effects. The real-world instrument effects such as the frequency-dependent and pointing-dependent beam shape, correlated non-Gaussian instrumental noise are so complicated that may affect the signal estimation. Besides that, additional complex foreground components such as extragalactic foregrounds, bright point sources, the thermal SZ effects and unknown dust components may significantly complicate the foreground removal as well. It is therefore important to test ABS in more realistic simulations. We will also apply ABS to the CMB polarization maps with hope of detecting B-modes in future work.

\section*{Acknowledgments}
This work was supported by the National Science Foundation of China (11653003, 11621303, 11433001, 11320101002, 11403071, 11773021), National Basic Research Program of China (2015CB85701), the National Key R\&D Program of China (2018YFA0404601). LS is supported by the National Natural Science Foundation of China (11603020, 11633001, 11173021, 11322324, 11421303, and 11653002), and by the project of Knowledge Innovation Program of Chinese Academy of Science, the Fundamental Research Funds for the Central Universities, and the Strategic Priority Research Program of the Chinese Academy of Sciences Grant No. XDB23010200.



\appendix
\section{Convergence test for shift parameter}

\begin{figure*}[htpb]
\centering
   \includegraphics[width=3.2in] {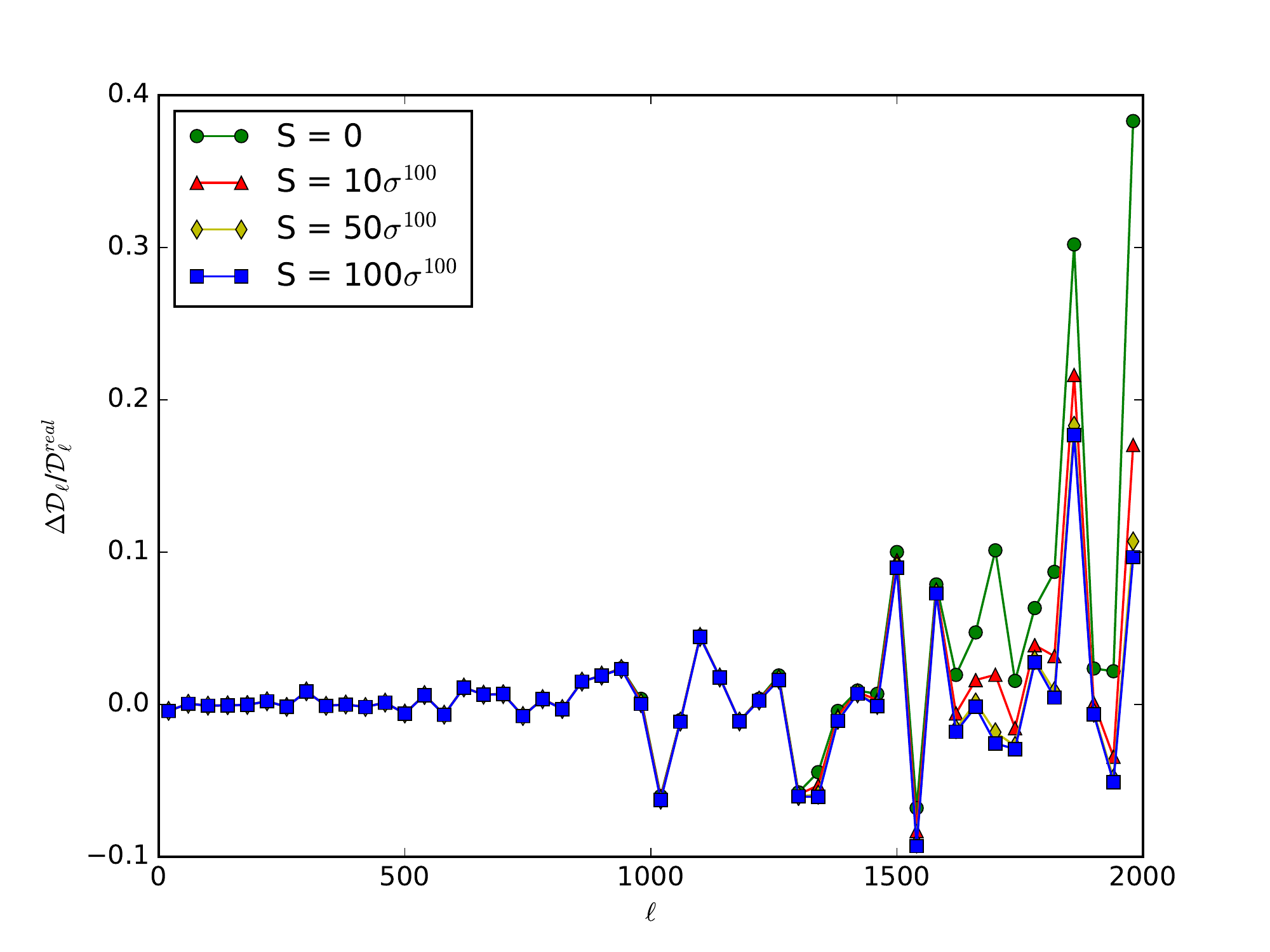}
\caption{\label{fig:ds} Results of the convergence test for the shift parameter $\mathcal{S}$.The noised level in each frequency has been manually {\it enhanced by a factor of 100}, dubbed as $\sigma^{100}$, to strengthen impacts of $\mathcal{S}$ in the recovered band powers $\mathcal{D}_\ell$. As seen, the derived band powers in the noise-dominant regime ($\ell\gtrsim 1200$) deviate strongly from the true one when $\mathcal{S}=0$, whereas the deviations gradually become smaller when increasing $\mathcal{S}$ and eventually converge to that of $\mathcal{S}=100\sigma^{100}$. The obtained band powers for $\mathcal{S} = 50\sigma^{100}$ and $100\sigma^{100}$ are essentially identical. Note that, the fiducial noise level $\sigma$, expected from the Planck, is so small that the relative changes from varying $\mathcal{S}$ from 0 to 100$\sigma$ are negligibly small (very similar to the signal-dominated regime at $\ell\lesssim 1200$  shown here), with the level of $<10^{-4}$ over all $\ell$ range.} 
\end{figure*}

The shift parameter $\mathcal{S}$ can play a major role in power spectrum estimation, especially for the regime where the signal-to-noise ratio (S/N) is low and less than unity. It is therefore very important to choose an appropriate  $\mathcal{S}$ for accurately and unbiasedly estimating the extremely faint primordial B-modes. Our simulations confirms that, the value of $\mathcal{S}$ has to be at least an order of magnitude greater than the noise level $\sigma$, so as to pass the convergence test. In Fig.~\ref{fig:ds}, we show the impact of $\mathcal{S}$ on the recovered band powers by varying $\mathcal{S}$, where the simulated noise level in each frequency has been increased by a factor of 10 manually. With this enhancement, the noise will dominate at the high-$\ell$ regime ($\ell\gtrsim1200$), and the recovered signals at such regime are changed with $\mathcal{S}$ rapidly. In comparison with the cases $\mathcal{S}=50$ and $100\sigma^{\sigma100}$, the results are fully converged to a relative change below 1\%  such that no measurable improvement can be expected within increasing $\mathcal{S}$. There are almost no differences in the signal-dominated regime at $\ell\lesssim1200$, with the relative changes smaller than $10^{-4}$ on average. The typical value of $\mathcal{S}$ is therefore chosen as about $50\sigma$ in this study to obtain reliable band powers, and any larger $\mathcal{S}$ will yield almost identical results.

\section{tests on ``full sky'' and ``two-times-stronger synchrotron'' maps}

\begin{figure*}[htbp]
\centering
\mbox{
 \subfigure[30GHz]{
   \includegraphics[width=2.3in] {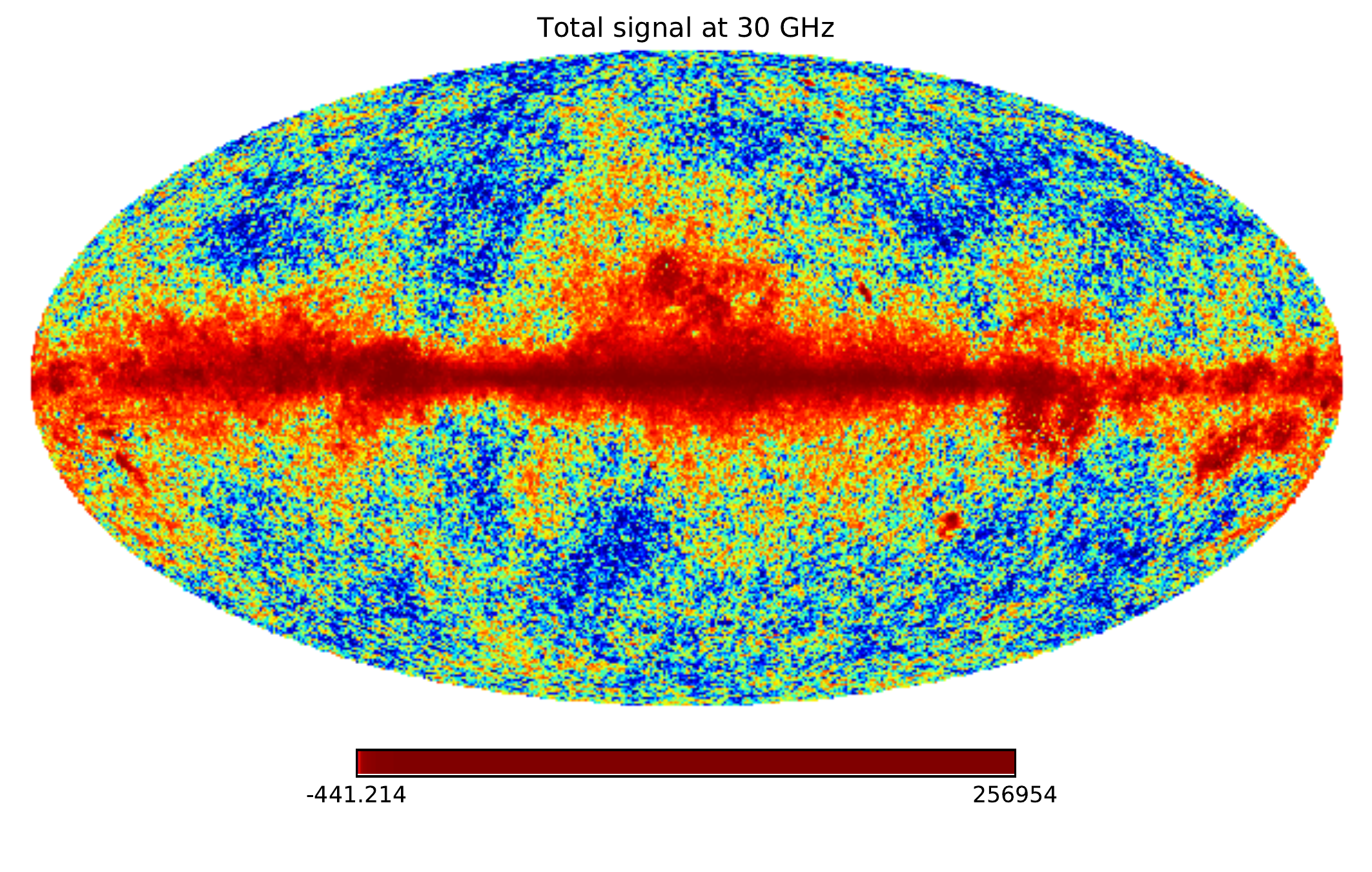}
    }
}

\mbox{
\subfigure[44GHz]{
   \includegraphics[width=2.3in] {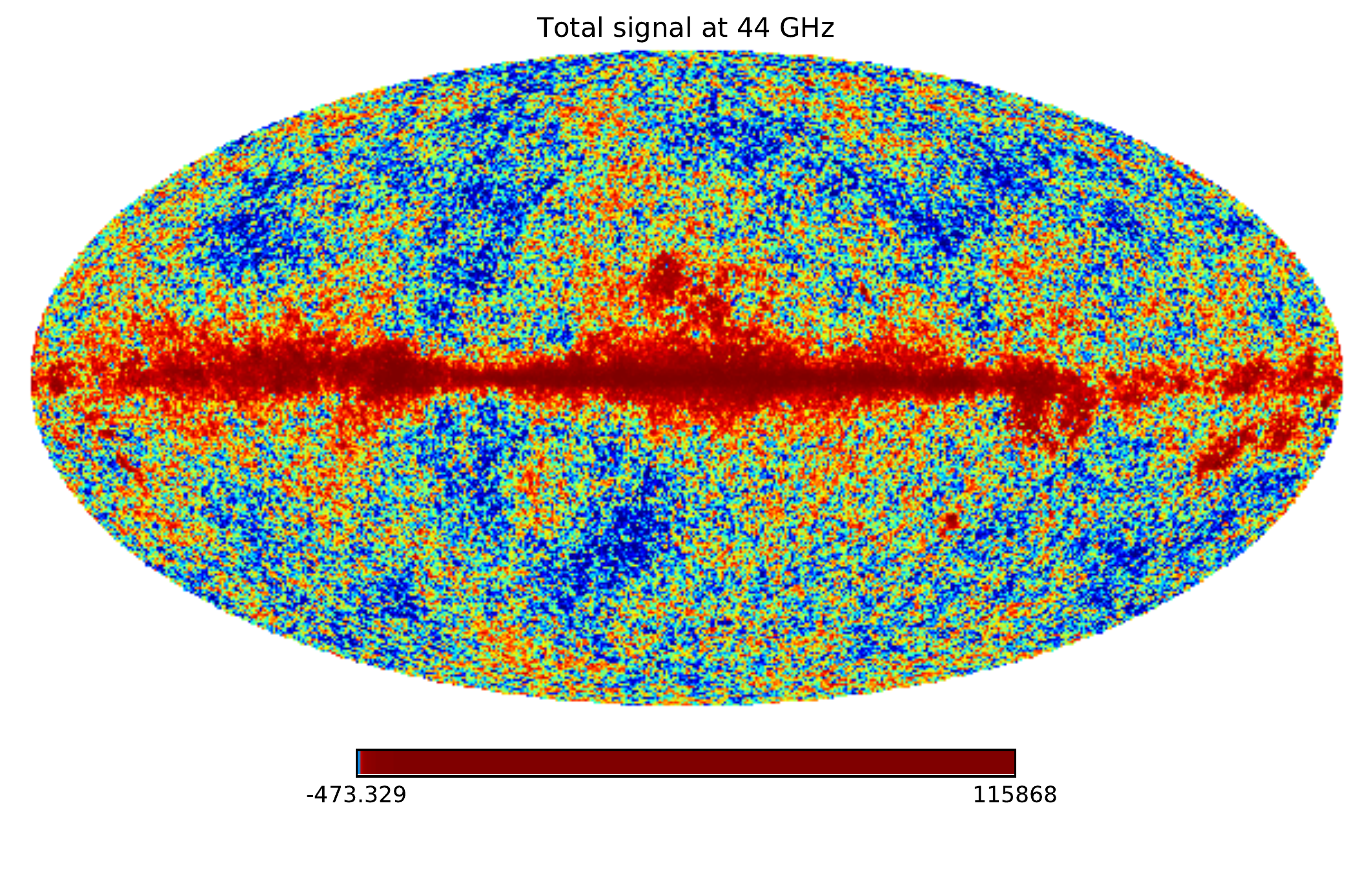}
   }

\subfigure[70GHz]{
   \includegraphics[width=2.3in] {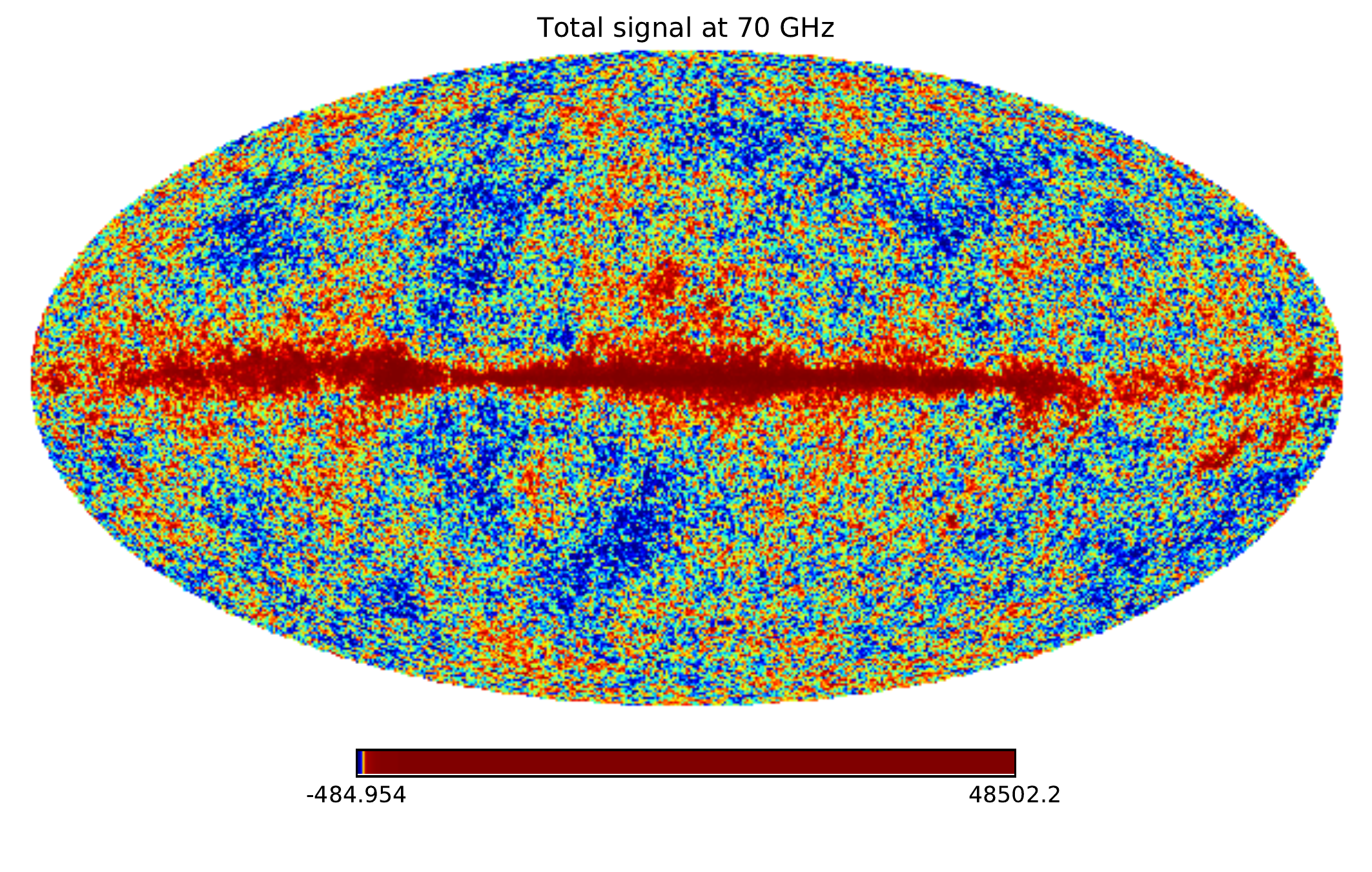}
   }

\subfigure[95GHz]{
   \includegraphics[width=2.3in] {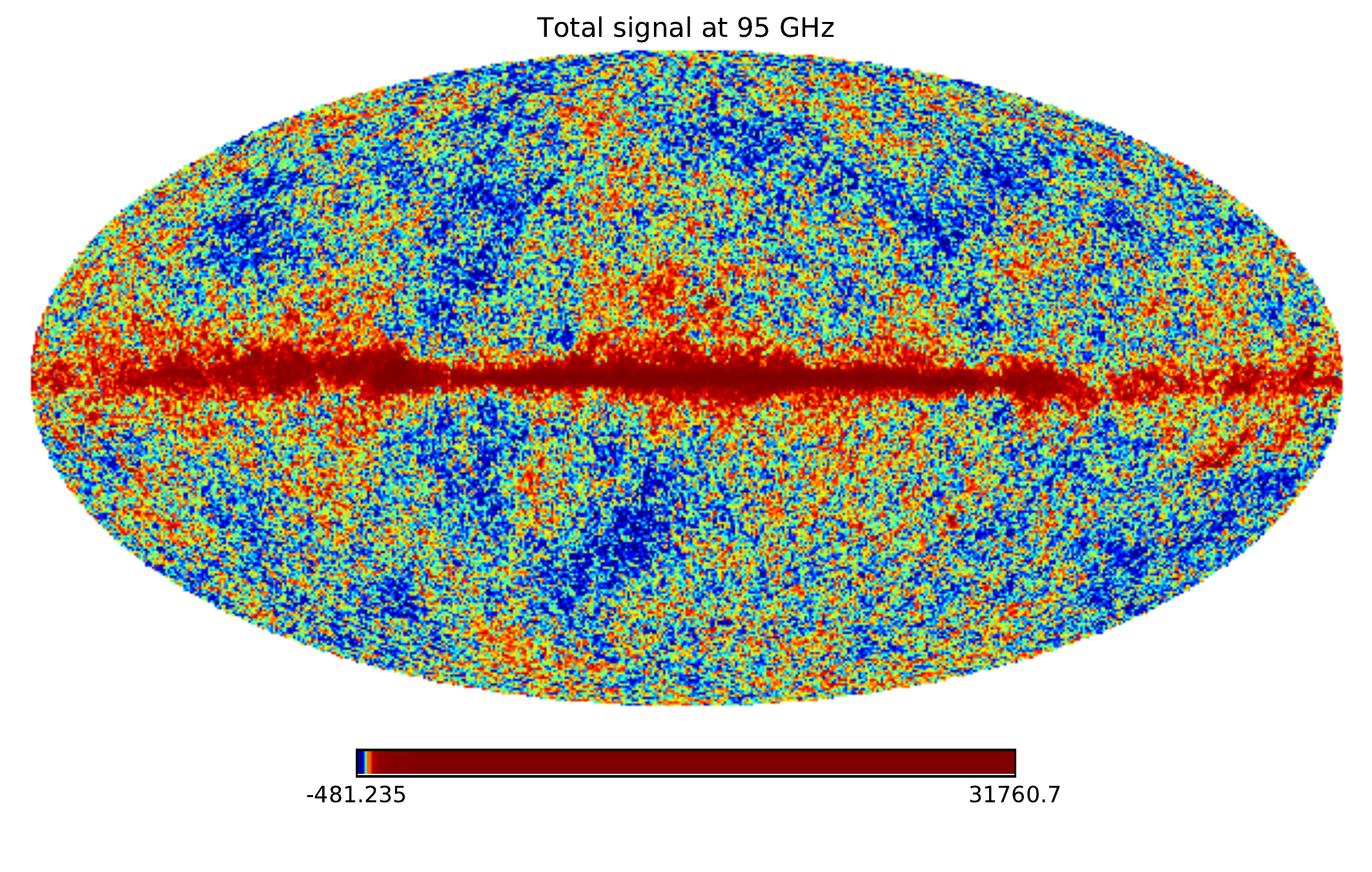}
   }
}

\mbox{
\subfigure[150GHz]{
   \includegraphics[width=2.3in] {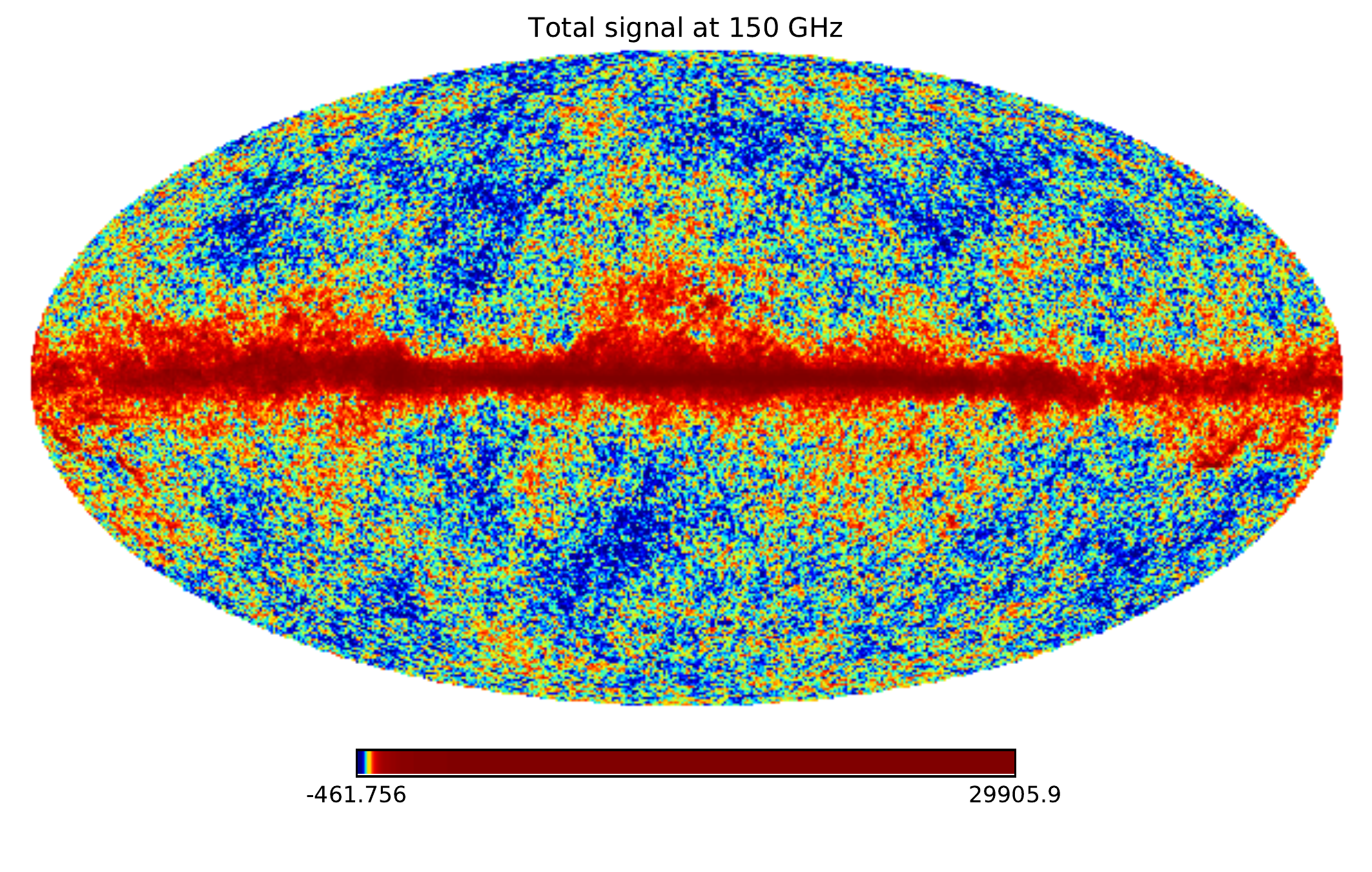}
   }

\subfigure[217GHz]{
   \includegraphics[width=2.3in] {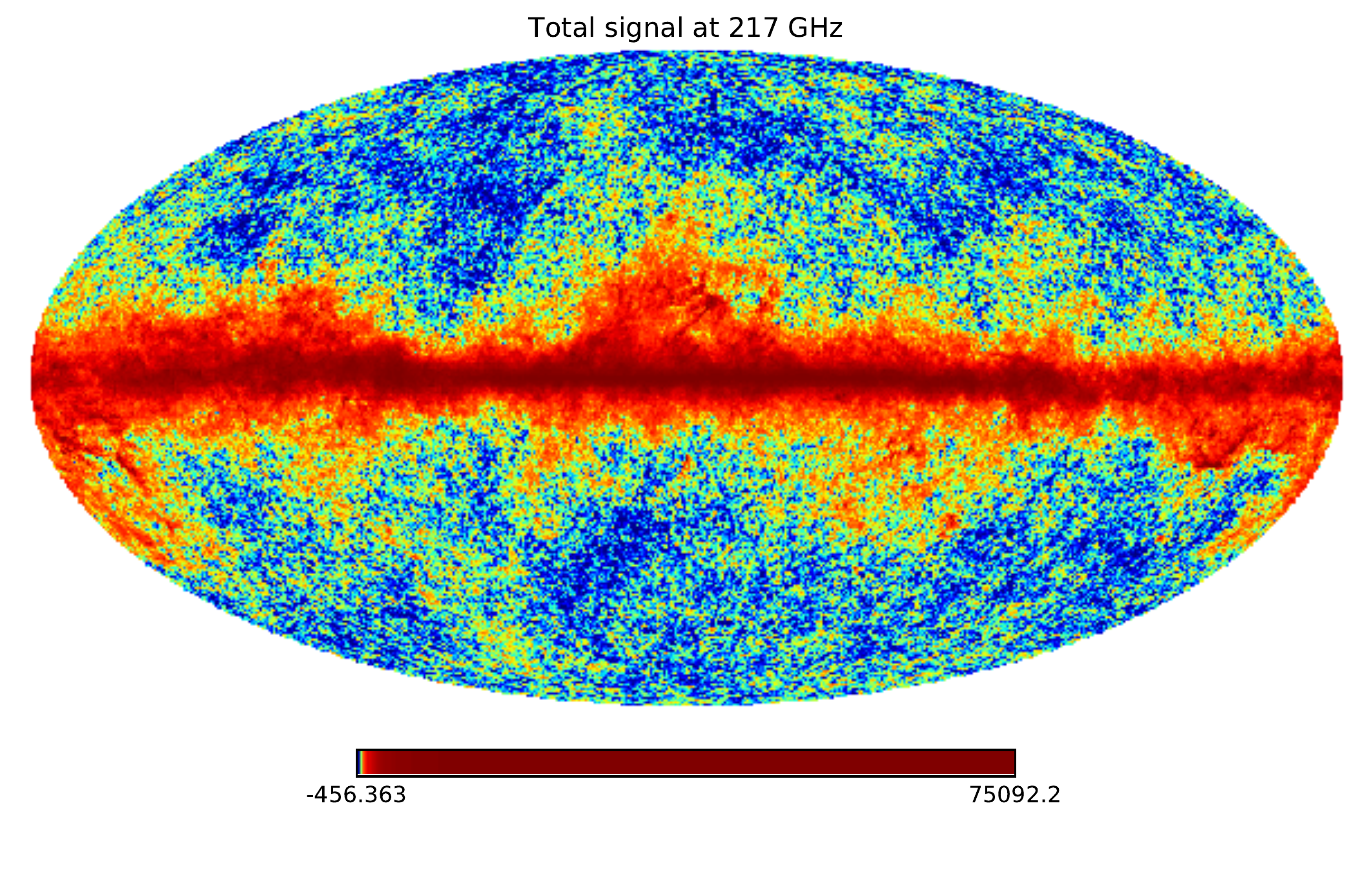}
   }

\subfigure[353GHz]{
   \includegraphics[width=2.3in] {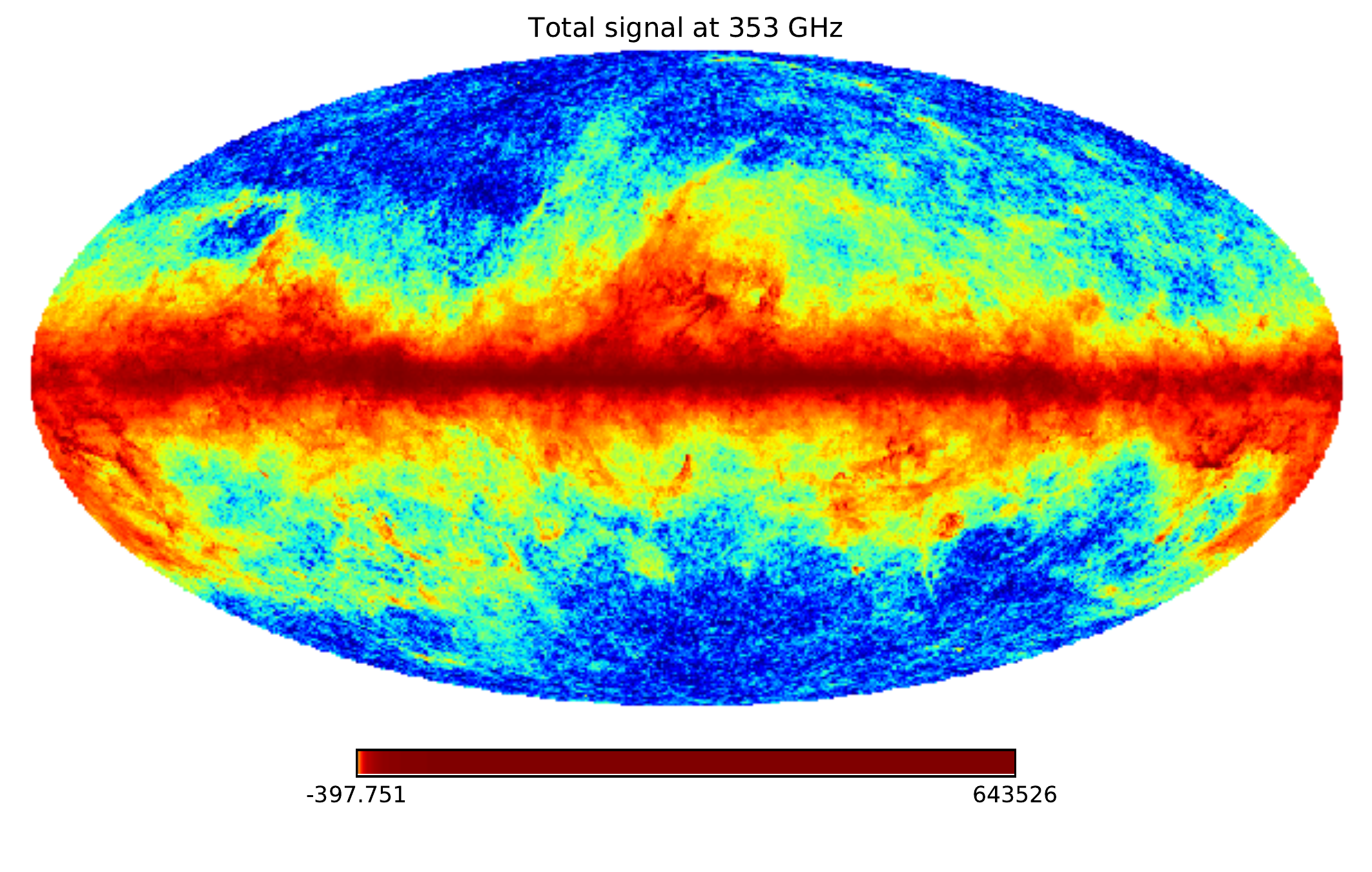}
   }
}

\caption{Same as Fig.~\ref{fig:map}, but for the case ``full sky'' at the entire frequency channels used for CMB band power estimation to further test the ABS approach.}
\label{fig:map-unmask}
\end{figure*}

To further assess and validate the reliability of the ABS approach in more extreme situations, we have repeated the ABS-based analyses using, however, two sets of frequency maps instead: ``full sky'' (shown in Fig.~\ref{fig:map-unmask}) and ``two-times-stronger synchrotron''. 
  
\begin{figure*}[htpb]
\centering
\mbox{
 \subfigure[30GHz]{
   \includegraphics[width=2.1in] {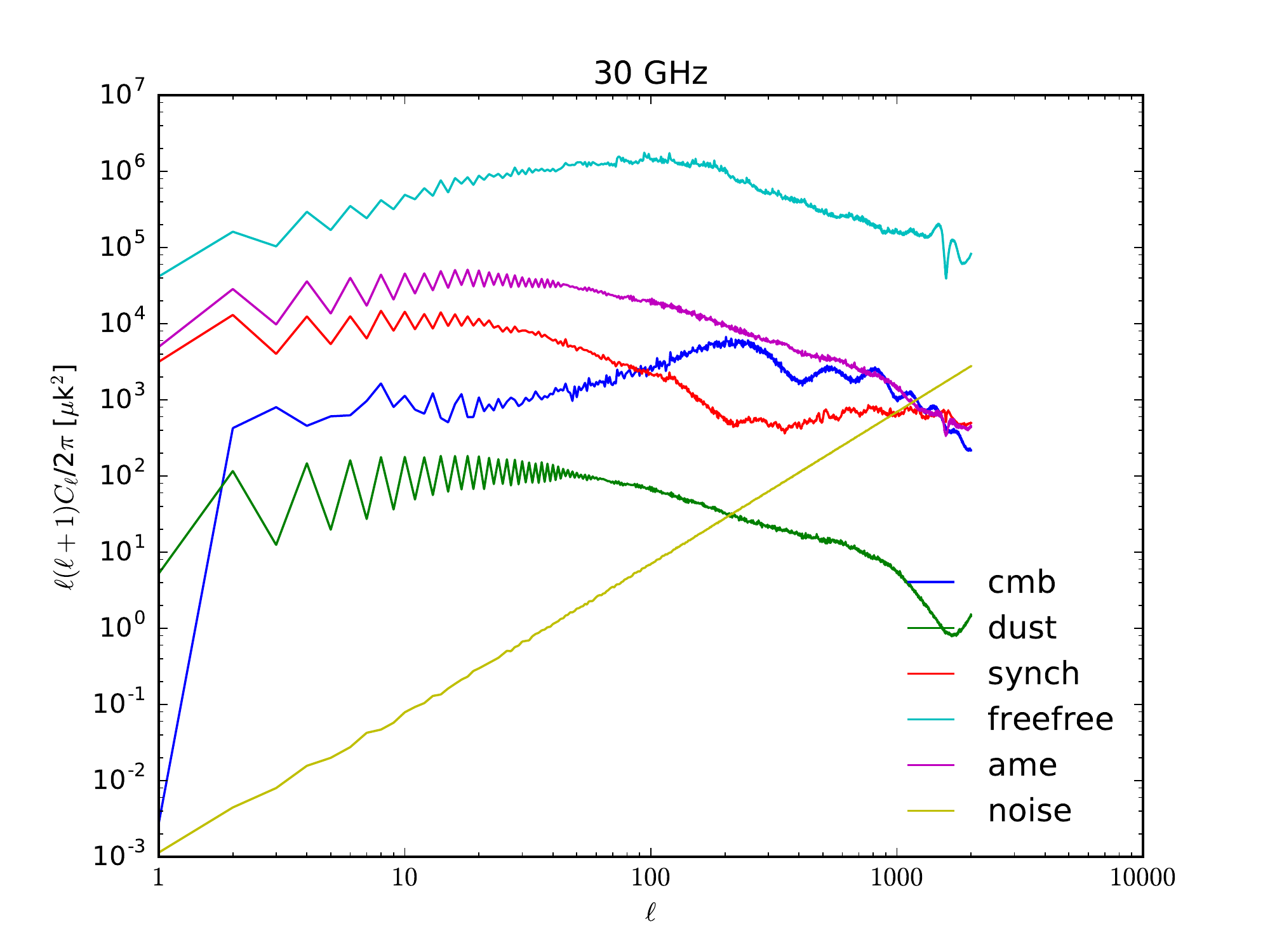}
    }
}

\mbox{
\subfigure[44GHz]{
   \includegraphics[width=2.1in] {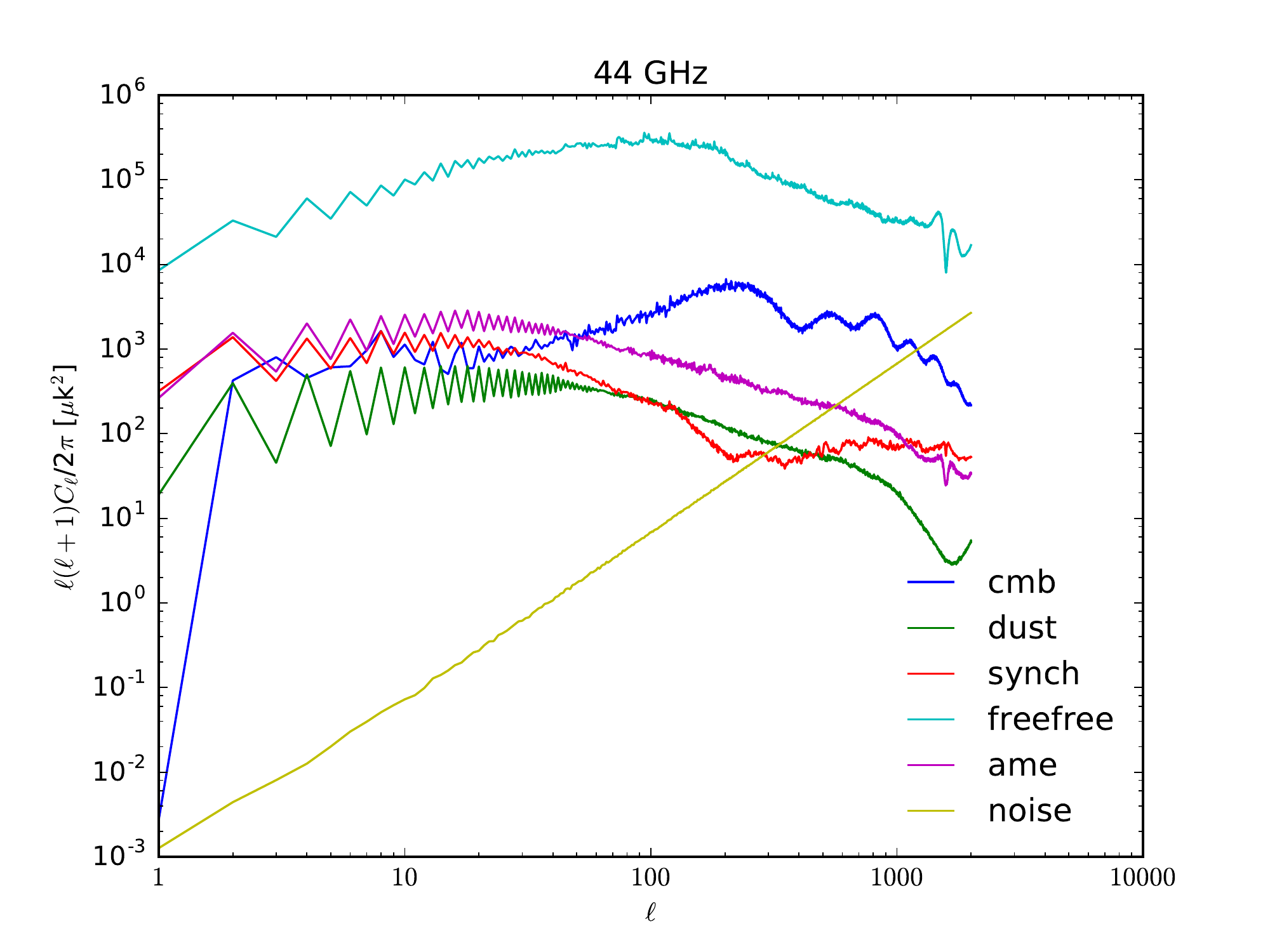}
   }

\subfigure[70GHz]{
   \includegraphics[width=2.1in] {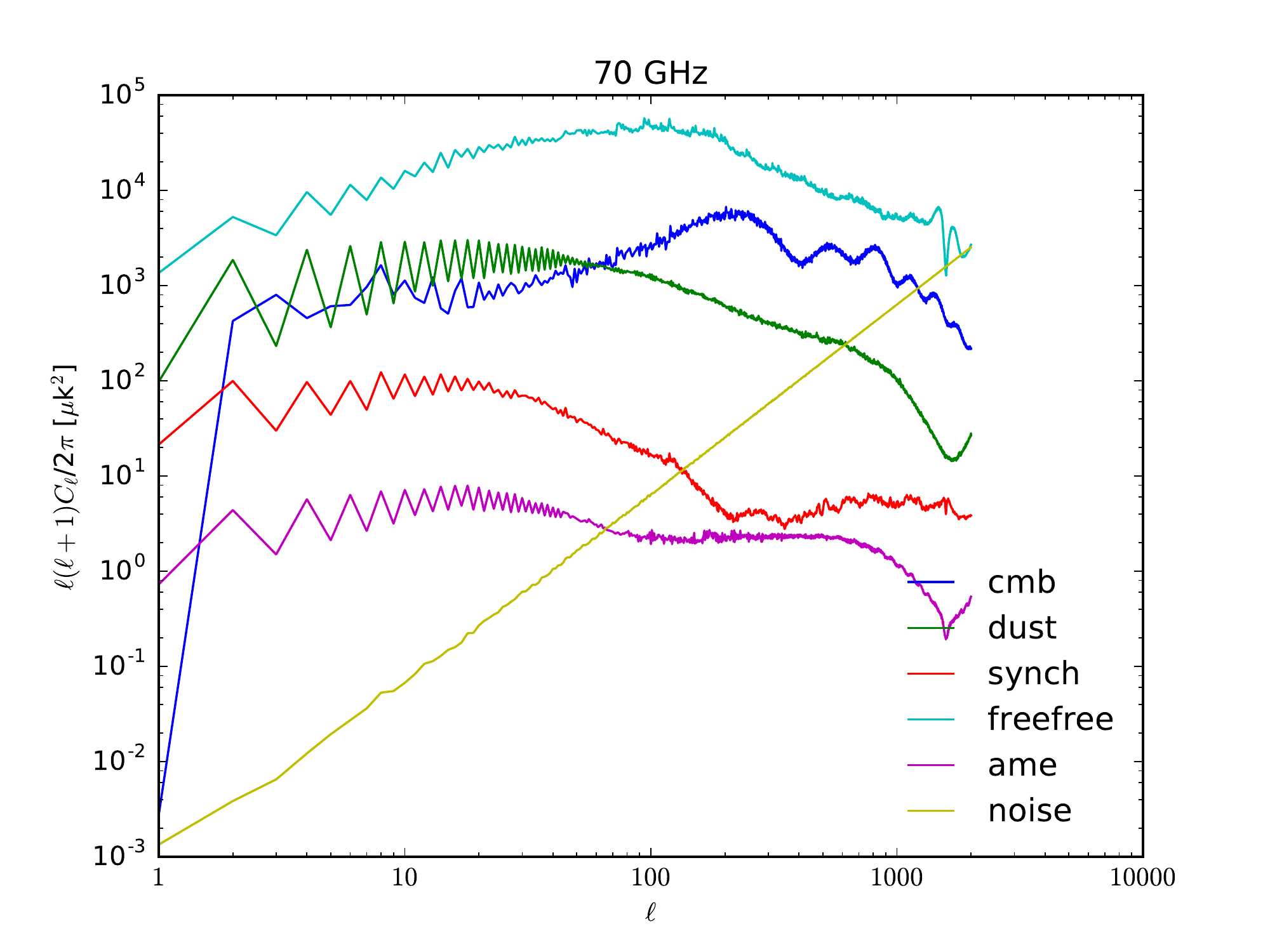}
   }

\subfigure[95GHz]{
   \includegraphics[width=2.1in] {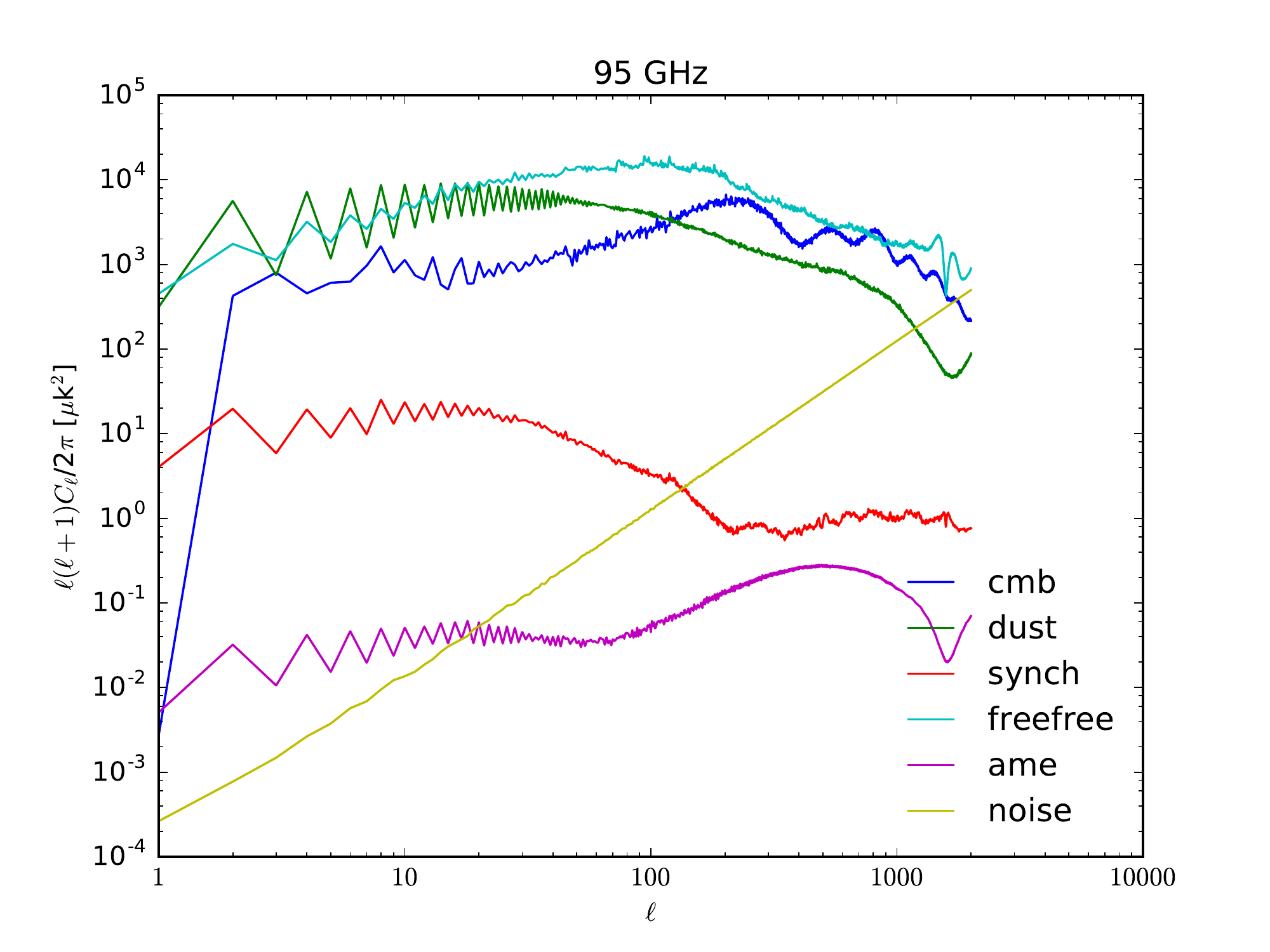}
   }
}

\mbox{
\subfigure[150GHz]{
   \includegraphics[width=2.1in] {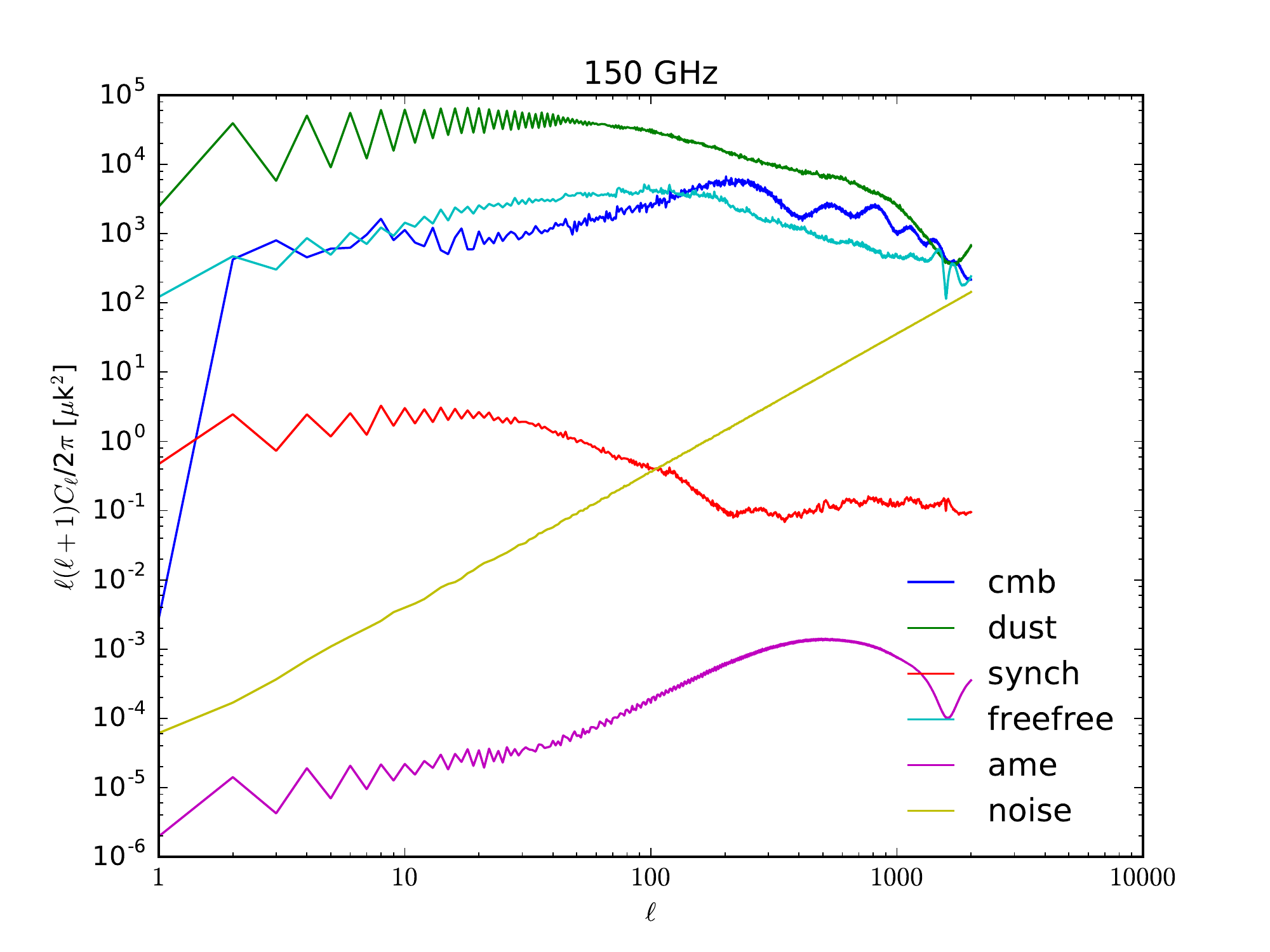}
   }

\subfigure[217GHz]{
   \includegraphics[width=2.1in] {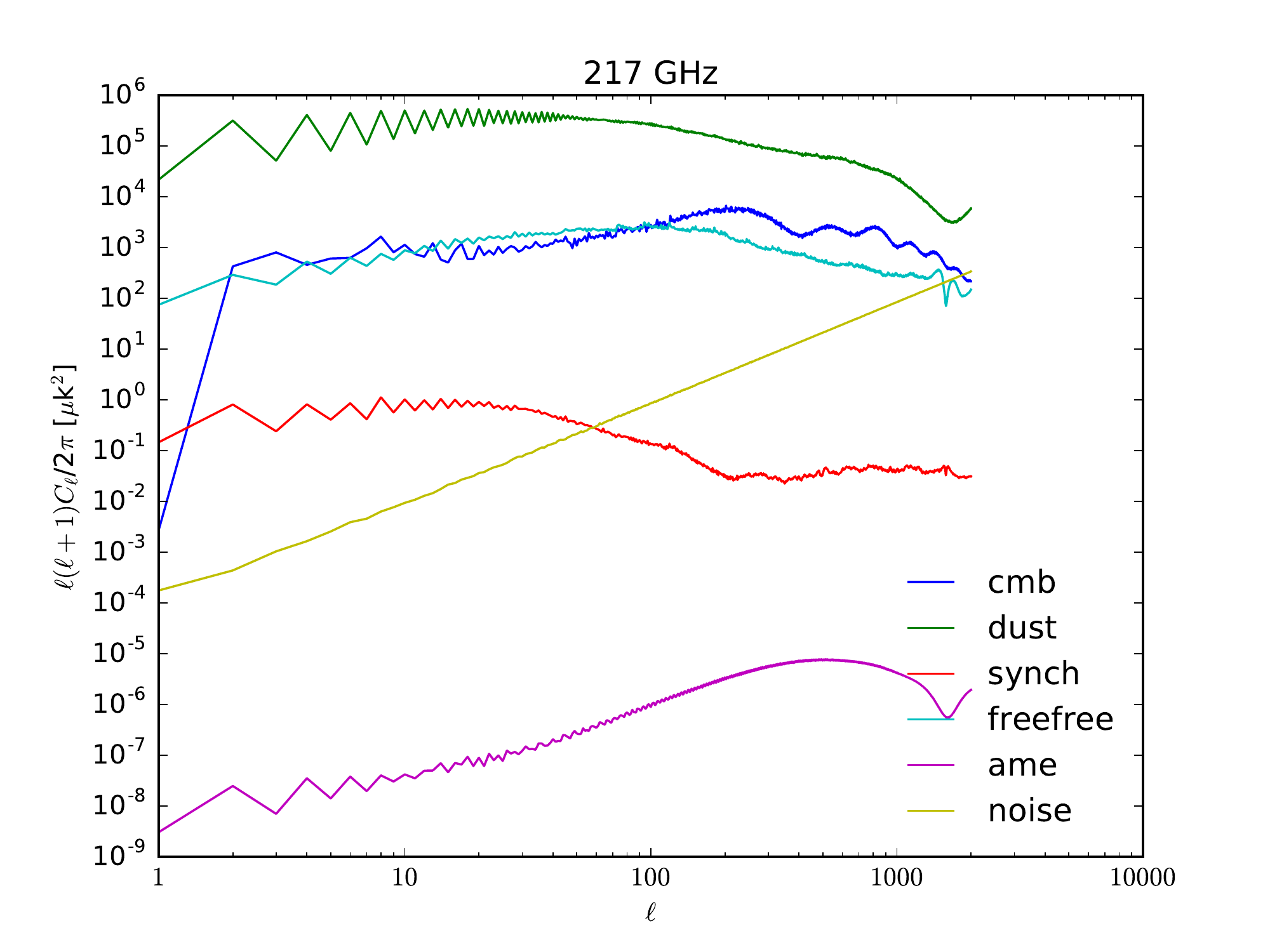}
   }

\subfigure[353GHz]{
   \includegraphics[width=2.1in] {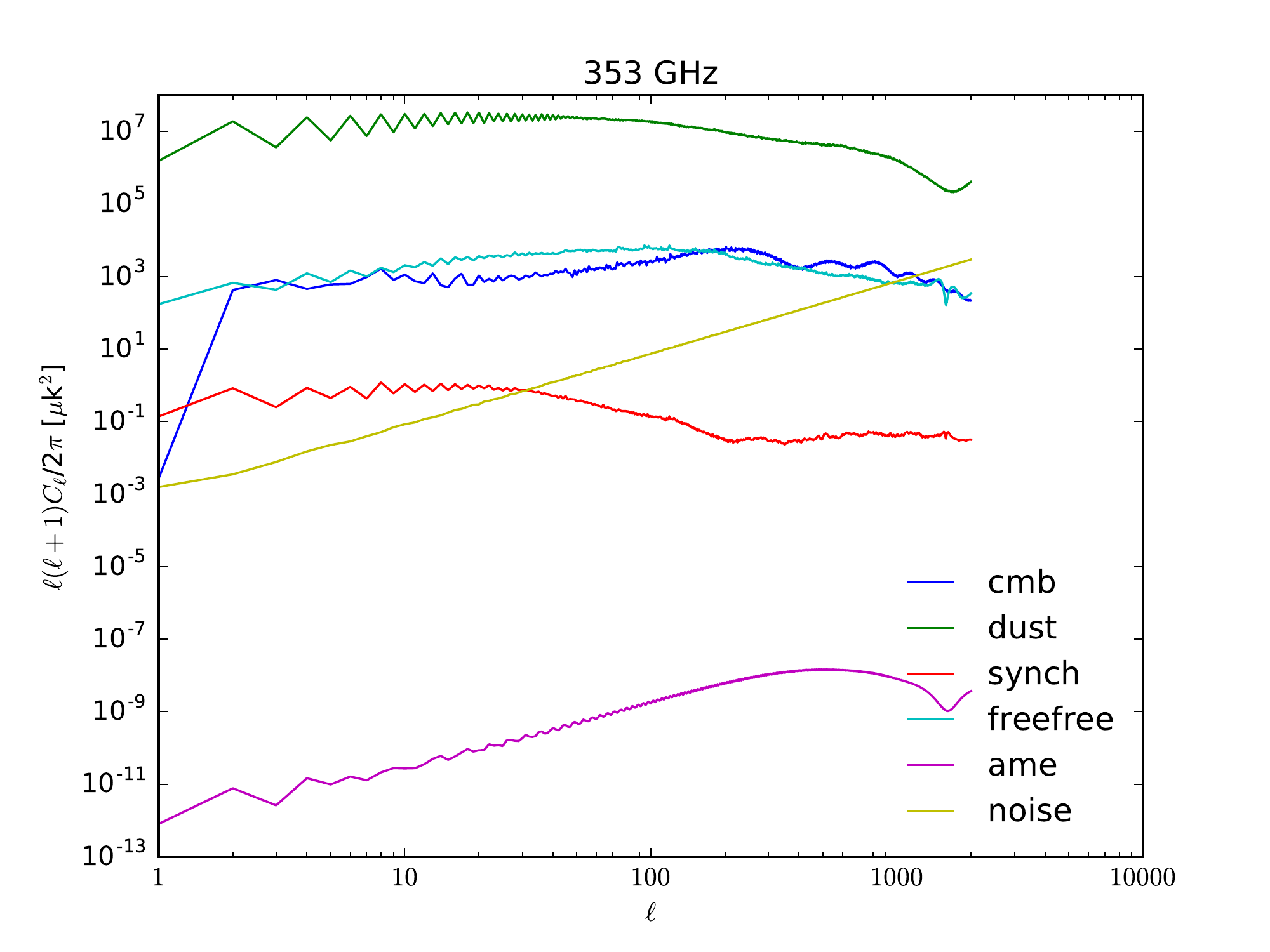}
   }
}

\caption{The corresponding power spectra of all microwave sources at the Planck maps in Fig.~\ref{fig:map-unmask} to clearly show the foreground and the noise contamination to CMB.}
\label{fig:cl-unmask}
\end{figure*}

The estimates of the CMB based on such sets of frequency maps are very similar to those obtained in the case ``Galactic-plane cut'' shown in Sect.~\ref{sect:test}. In the case ``full sky'' where the full sky seven-frequency maps are used in the CMB power spectrum estimation, even though the powers of all foreground components become an order of magnitude higher than those in the case ``Galactic-plane cut''  (compared with Fig.~\ref{fig:cl-unmask} and the lower panels of Fig.~\ref{fig:map} ), the recovered spectrum is still unbiased, well within the 2-$\sigma$ error bars at most $\ell$ bins, and the relative errors of the recovered power spectrum now is below 1\% which is slightly larger than those from the case ``Galactic-plane cut''.

As a consistency check, one may wonder whether the derived deviation and associated error in the ``full sky'' case (denoted by ``C'') are consistent with the results from the ``Galatic-plane cut'' (``A'') and the ``inside Galactic-plane'' (``B'') cases. As expected, the relative deviation in the ``full sky'' case, $r_{\rm C}$, can be approximated by $r_{\rm C} =  \Delta \mathcal{D}_{\rm C}/\mathcal{D}^{\rm real}_{\rm C}\approx \Delta \mathcal{D}_{\rm C}/\left(\mathcal{D}^{\rm real}_{\rm A}  +\mathcal{D}^{\rm real}_{\rm B}\right)\approx \left(r_{\rm A}\mathcal{D}^{\rm real}_{\rm A}+ r_{\rm B}\mathcal{D}^{\rm real}_{\rm B}\right)/\left(\mathcal{D}^{\rm real}_{\rm A}  +\mathcal{D}^{\rm real}_{\rm B}\right)$. Inserting the derived values for $r_{\{\rm A,B,C\}}$ and $\mathcal{D}^{\rm real}_{\{\rm A, B, C\}}$ from the simulations, we find this approximation is roughly valid. For the statistical error in the power spectrum, the relation among these three cases in terms of the sky coverage $f_{\rm sky}$, $\sigma_{\rm C} \simeq  f^{\rm A}_{\rm sky}\sigma_{\rm A}+ f^{\rm B}_{\rm sky}\sigma_{\rm B}$, is also valid, as confirmed by the simulations. Furthermore, the ratio of the systematic error (i.e. deviation) and the statistical error in case C is somewhat greater than either that in case A and B, unlike one might expect that the ratio in case C should be in between the case A and B. It may be caused by the effects from some nonlinear processes (such as the ``cut off'' in eigenmodes) in ABS.

\begin{figure*}[htpb]
\centering
\mbox{
 \subfigure[Full sky]{
   \includegraphics[width=3.2in] {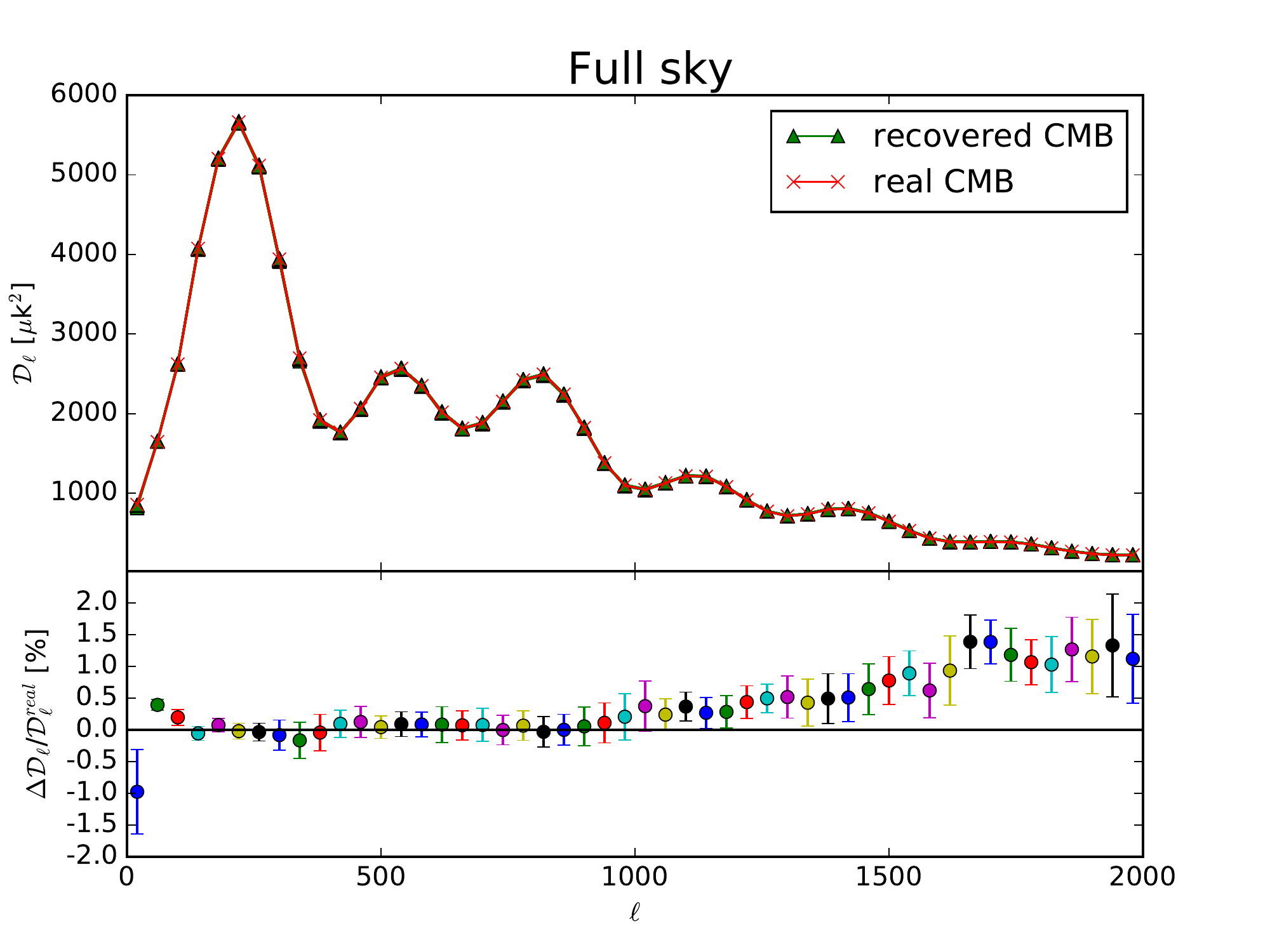}\label{fig:unmask}
    }

\subfigure[two-times-stronger synchrotron]{
   \includegraphics[width=3.2in] {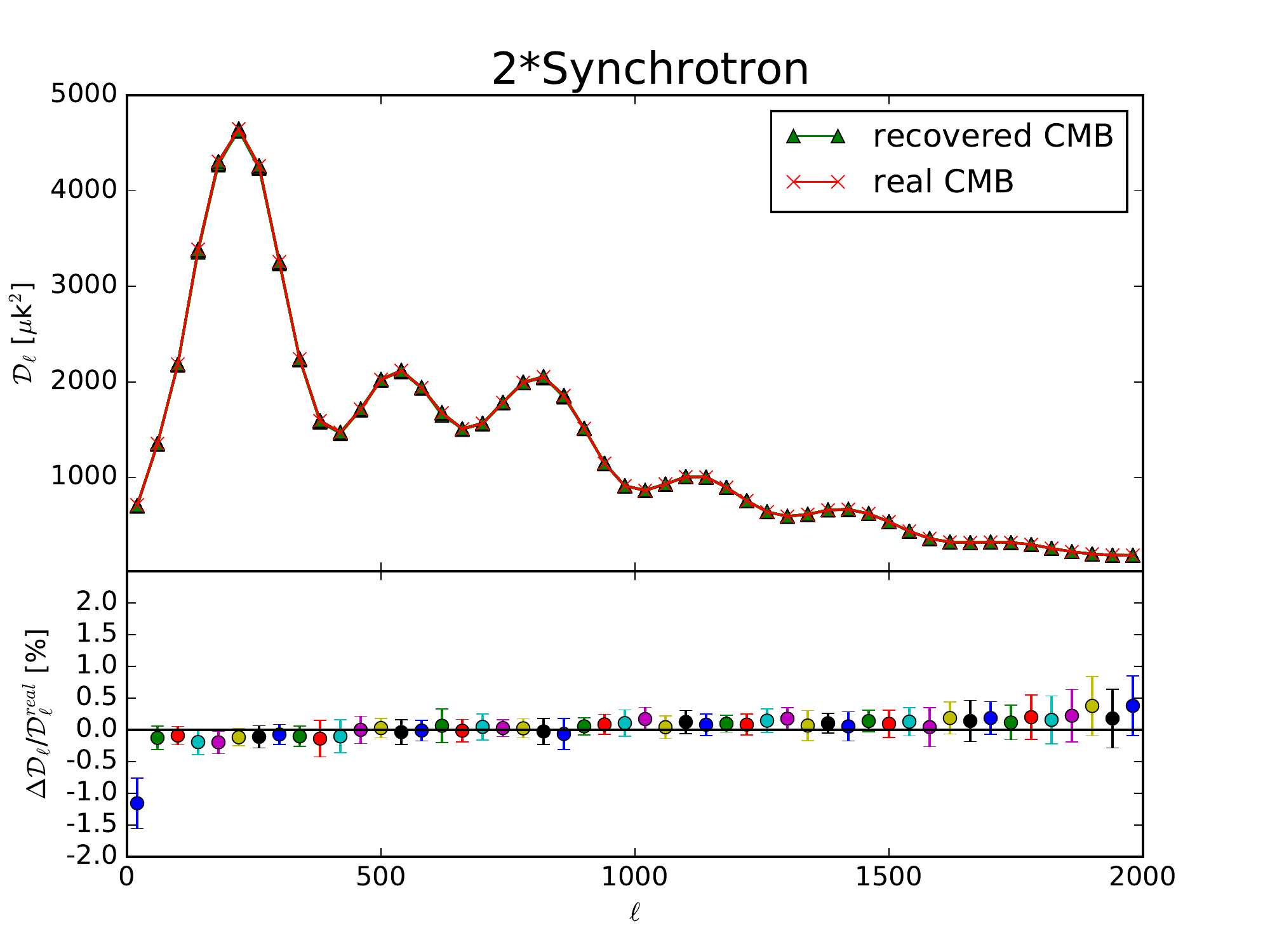}\label{fig:syn2}
   }

}

\caption{Same as Fig.~\ref{fig:rec-mask}, but for the cases of ``full sky'' ({\it left panel}) and ``two-times-stronger synchrotron'' ({\it right panel}).}
\label{fig:rec-2}
\end{figure*}

Next, we test the ABS approach by varying the amplitude of foreground. In the case ``two-times-stronger synchrotron'', we increase the amplitude of the synchrotron foreground map at each frequency by a factor of two (much larger than the uncertainty in its modeling) and keep the other components fixed, together with our fiducial mask removing the region within $10^\circ$ from the Galactic plane. The results are demonstrated in Fig.~\ref{fig:syn2}, clearly implying that the accuracy of the recovery remains unchanged with respect to Fig.~\ref{fig:rec-mask}. The above results of these two cases confirm that the ABS-based CMB estimation is very flexible and it cannot be significantly biased by the much stronger foreground contamination. Therefore, we can conclude again that the ABS approach provide a robust way and an unbiased estimator that enable us to extract the underlying CMB power spectrum from observed frequency maps accurately.

\end{document}